\documentclass{article}

\usepackage{authblk}
\usepackage[english]{babel}
\usepackage{microtype}      
\usepackage{multirow}      
\usepackage[colorlinks=true, allcolors=blue]{hyperref}
\usepackage{amsfonts}
\usepackage{url}
\usepackage{listings}
\usepackage{makecell}
\usepackage{booktabs}
\usepackage{wrapfig}
\usepackage{algorithm}
\usepackage{algpseudocode}
\usepackage{bbm}
\usepackage[utf8]{inputenc} 
\usepackage[T1]{fontenc}    
\usepackage{booktabs}       
\usepackage{amsfonts}       
\usepackage{nicefrac}       
\usepackage{microtype}      
\usepackage{xcolor}         
\usepackage{mathtools}
\usepackage{float}
\usepackage{amsthm}
\usepackage{enumitem}
\usepackage{subcaption}
\usepackage[letterpaper,top=2cm,bottom=2cm,left=3cm,right=3cm,marginparwidth=1.75cm]{geometry}
\lstdefinestyle{promptstyle}{
    backgroundcolor=\color{gray!10},
    basicstyle=\ttfamily\fontsize{7}{9}\selectfont,
    frame=single,
    breaklines=true
}

\usepackage[round]{natbib}
\usepackage{amsmath}
\usepackage{graphicx}
\usepackage[colorlinks=true, allcolors=blue]{hyperref}
 
\DeclarePairedDelimiterX\set[1]\lbrace\rbrace{#1}

\title{RLRF: Competitive Search Agent Design via Reinforcement Learning from Ranker Feedback}

\author[1]{Tommy Mordo}
\author[2]{Sagie Dekel}
\author[2]{Omer Madmon}
\author[1]{Moshe Tennenholtz}
\author[1]{Oren Kurland}

\affil[1]{Faculty of Data and Decision Sciences\\ Technion -- Israel Institute of Technology}
\affil[ ]{\texttt{\{tommymordo, moshet, kurland\}@technion.ac.il}}

\affil[2]{Faculty of Data and Decision Sciences\\ Technion -- Israel Institute of Technology}
\affil[ ]{\texttt{\{sagie.dekel, omermadmon\}@campus.technion.ac.il}}

\begin{document}
\date{}
\maketitle
\begin{abstract}
Competitive search is a setting where document publishers modify them to improve their ranking in response to a query. Recently, publishers have increasingly leveraged LLMs to generate and modify competitive content. We introduce Reinforcement Learning from Ranker Feedback (RLRF), a framework that trains LLMs using preference datasets derived from ranking competitions. The goal of a publisher (LLM-based) agent is to optimize content for improved ranking while accounting for the strategies of competing agents. We generate the datasets using approaches that do not rely on human-authored data. We show that our proposed agents consistently and substantially outperform previously suggested approaches for LLM-based competitive document modification.  We further show that our agents are effective with ranking functions they were not trained for (i.e., out of distribution) and they adapt to strategic opponents. These findings provide support to the significant potential of using reinforcement learning in competitive search.
\end{abstract}

\newcommand{\staticDocsGeneration}{SG}
\newcommand{\dynamicDocsGeneration}{DG}
\newcommand{\baseLLM}{base LLM}
\newcommand{\testEnv}{LEMSS}
\newcommand{\nivList}{LSW}
\newcommand{\nivPair}{PAW}
\newcommand{\listwisePrompt}{listwise prompt}
\newcommand{\pairwisePrompt}{pairwise prompt}
\newcommand{\noFeedbackPrompt}{no-feedback prompt}
\newcommand{\player}{player}
\newcommand{\game}{game}
\newcommand{\staticAgent}{SA}
\newcommand{\dynamicAgentPairwise}{DPA}
\newcommand{\dynamicAgentListwise}{DLA}
\newcommand{\winRateMetric}{win rate}
\newcommand{\WiningHomogeneityMetric}{wining homogeneity}
\newcommand{\homoEval}{Ho}
\newcommand{\hetroEval}{He}
\newcommand\todo[1]{\textcolor{red}{#1}}
\newcommand{\gpt}{gpt-4}
\newcommand{\rl}{RA agent}
\newcommand{\prompt}{NA agent}

\section{Introduction}

\emph{Competitive Search} refers to a search setting where strategic document authors actively optimize their documents' content to improve ranking in response to a query induced by a search engine \citep{compSearch}. 
Ranking competitions are particularly intense in commercial domains, where a higher search rank directly translates into increased traffic, influence, and revenue \citep{joachims2017accurately}. As search algorithms evolve, so do the modifications applied by publishers, making competitive search a dynamic interplay between the search algorithms and strategic content creation.

While traditional publishers' strategies often relied on surface-level techniques such as keyword stuffing (designed to exploit the bag-of-words nature of early search algorithms; \citealp{zuze2013keyword,drivas2017stuffing}) or non-content-based approaches (aimed at manipulating PageRank-based systems; \citealp{alice2006manipulability,bar2007manipulating}), the rise of large language models (LLMs) has fundamentally reshaped the competitive search landscape. Modern search engines increasingly rely on advanced neural ranking methods such as dense retrieval\footnote{Dense retrieval refers to a retrieval paradigm in which both queries and documents are encoded into dense vector representations (typically using neural networks), and relevance is estimated via vector similarity (e.g., dot product), rather than sparse term overlap as in traditional methods \citep{zhao2024dense}.}, which prioritize semantic understanding over exact keyword matches \citep{zhao2024dense}. As a result, publishers now focus on crafting content that aligns with the deeper meaning and intent behind user queries. 

At the same time, the rise of LLMs has made it easier for publishers to engage in this new form of semantically driven optimization. LLMs not only excel in core natural language processing tasks such as sentiment analysis and text generation \citep{brown2020language_lang1,peng2023towards_lang3, zhang2023extractive_lang2, susnjak2023applying_lang4,wang2023chatgpt_lang5}, but also in competitive tasks that require strategic reasoning \citep{shapira2024glee,raman2024steer,akata2025playing}, positioning them as powerful tools for navigating the increasingly complex and competitive search ecosystem. This dual role of LLMs, as both a force in ranking algorithms and a content creation tool for publishers, has created a new era of competitive search centered on strategic content design \citep{nachimovsky2025multi}.

The role of LLM-based agents as strategic publishers in competitive search environments has not yet been systematically studied. \citet{mordo_lemss_2025} introduced a simulation framework that models ranking competitions involving both human and LLM-based participants. \citet{bardas_automatic_2025} employed a framework to evaluate the effectiveness of LLM agents in one-shot competitive search settings under different prompting and feedback strategies. This raises a natural question: \textbf{Can LLM-based strategic agents be improved beyond prompting by training them — using reinforcement learning — to optimize for ranking competition objectives, i.e., to be ranked as highly as possible during the competition?}



In this work, we introduce a novel paradigm for training LLM-based agents in competitive search environment that leverages reinforcement learning (RL) alignment techniques to improve the content produced by agents in terms of rankings. The key idea is to align the LLMs using feedback induced from the ranker's output (i.e., the ranking), where this feedback is reformulated as prompts for the LLM-based agents. By incorporating this feedback, the agent learns to produce content that is more likely to be ranked higher across a variety of queries and competitive contexts. Importantly, the RL-based alignment occurs only at training time; at test time, the agents operate solely through prompting, without additional optimization. We refer to this approach as \emph{Reinforcement Learning from Ranker Feedback (RLRF)}. Agents trained using this paradigm are henceforth referred to as RL-aligned agents or \textbf{RA agents} in short.
Our contributions are as follows:

\begin{itemize}[noitemsep, topsep=-5pt, leftmargin=*]
    \item We formalize the setting of \textit{competitive search} as a learning problem in which LLM-based agents generate content to maximize their rank in a dynamic ranking environment.
    \item We introduce the novel RLRF methodology, which aligns the LLM with the competitive ranking objective. We characterize two key aspects of the learning process of \rl{}s: (i) aligning with the search engine’s ranking function, and (ii) adapting to strategic opponents in a ranking competition.
    \item We train our agent on synthetic datasets generated using two approaches: Static Generation (SG), which produces documents' modifications independent of other agents, and Dynamic Generation (DG), which simulates multi-agent competition.
    \item We demonstrate the effectiveness of RLRF through extensive experiments in a controlled competitive search framework, showing that agents aligned with RLRF consistently outperform baseline prompting-based approaches across a range of queries and competitive settings.
    \item We show that \rl{}s trained with one ranker can transfer effectively to different ranking functions.  
\end{itemize}
\section{Related Work}

\paragraph{Game-theoretic Foundations of Competitive Search}
There is a growing body of work on competitive search settings where
document authors modify their documents so as to improve their future
ranking in response to queries \citep{compSearch}. Specifically, game
theoretic approaches were used, alongside empirical studies, to
analyze ranking paradigms
\citep{compSearch,ben2015probability,basat2017game,ben2019convergence,nachimovsky2024ranking,mordo2025searchresultsdiversificationcompetitive}
(e.g., whether they lead to equilibrium), to study authors' document
modification strategies \citep{raifer2017information,ben2019convergence,madmon2025nrd,madmon2025search},
and to explore potential corpus-based enrichment approaches to ensure
equilibrium \citep{nachimovsky2025power}. In contrast, we focus on
RL-based training of LLM agents that act as document authors.

\paragraph{LLMs in Competitive Environments}
LLMs have recently shown strong potential as rational agents in strategic interactions \citep{xi2023rise,fu2023improving,wang_survey_2024,guo2024economics,guo_large_2024,akata2025playing,xie2025strategic}. Recent benchmarks were used to evaluate LLM performance in complex multi-agent decision-making tasks, assessing both individual rationality \citep{raman2024steer,raman2025steer} and collective economic measures such as efficiency and fairness \citep{shapira2024glee}. One of the promising directions is simulating competitive tasks using LLMs \citep{zhao_competeai_2024}; the theoretical aspects are sometimes analyzed using game theoretic models \citep{mao2024alympicsllmagentsmeet}. As highlighted by \cite{nachimovsky2025multi}, LLMs can play different roles in the competitive search ecosystem. While most of the previous work focused on the ranker's perspective \citep{gao2024llm,wang2024towards,rathee2025guiding,guo2025mcranker}, we focus on utilizing LLMs to generate documents from the perspective of the (strategic) publisher.

\cite{bardas_automatic_2025} initiated the study of LLM-based agents, showing that few-shot LLMs can perform on par with human publishers in a single-round ranking promotion setting. In contrast, our work addresses a more complex and practical framework where agents modify their content across long-term interactions with other agents. Building on the competitive search simulation framework of \cite{mordo_lemss_2025}, we show that RLRF techniques can enhance LLM-based agents to outperform the few-shot agents of \cite{bardas_automatic_2025}.

\paragraph{RL in Competitive Settings}
RL has long been used to train agents in competitive and multi-agent environments, achieving remarkable success in board and video games \citep{vinyals2017starcraft,xenou2018deep,vinyals2019grandmaster,li2024fightladder}. More recently, RL from human feedback (RLHF) has emerged as a key technique for aligning large language models (LLMs) with human preferences in non-strategic tasks such as summarization and dialogue generation \citep{christiano2017deep,ouyang2022training,shen2023large,gao_towards_2024, tennenholtz_embedding-aligned_2024}. To scale this approach, RL from AI feedback (RLAIF) has been proposed, replacing human evaluators with LLM-based feedback to improve scalability \citep{bai2022constitutional,lee2024rlaifvsrlhfscaling}. Subsequent work applied RL-based techniques to enhance the decision-making abilities of LLMs \citep{unknown} and to optimize content generation in competitive landscapes \citep{sharma2022leveraging,coppolillo2024engagement}. RL has also been applied to recommendation systems to improve recommendation performance by optimizing long-term user engagement \citep{sun_rlrf4rec_2024}. More recently, \citet{10.1145/3726302.3730026} introduced an RL-based generator agent that strategically uploads items into recommender environments. While both works use LLM-based agents to generate content, their focus is on simulating generators to evaluate recommender systems and on aligning synthetic data with real-world distributions (e.g., YouTube). In contrast, our goal is to design long-term strategies for agents in multi-agent settings rather than to evaluate recommenders.

\paragraph{RL in Information Retrieval}
An RL-based relevance feedback approach improved retrieval effectiveness by
iteratively adapting to user interactions \citep{RLRelevance2020} (a.k.a., dynamic retrieval \citep{yang_dynamic_2016}).
RL was also used with LLMs to guide interaction with search engines \citep{jin_search-r1_2025} and to enhance query generation and expansion \citep{jiang2025deepretrieval,yang_aligned_2025}. In contrast to this line of work which focuses on the ranker, our focus is on content creation by publishers aiming to improve the ranking of their documents.

%
%
%
%
%
\section{Task Definition and Approach}\label{sec:task}
We address the task of designing a document authoring agent which competes in a repeated ranking game \citep{compSearch}. In each game, a fixed set of agents repeatedly compete for the highest ranking induced by an undisclosed ranking function for a given query. A competition consists of multiple games, where each game is associated with a distinct query. Each game lasts for several rounds. At the beginning of a game, each agent is assigned with an identical initial document. From the second round onward, all agents simultaneously modify their documents based on the ranking in the previous rounds. After all agents submit the modified versions of their documents, the system applies a non-disclosed ranking function; specifically, only the ordering of documents is provided every round. The goal of each agent is to strategically adapt its document over the course of a game in order to consistently achieve high ranks. A schematic illustration of a single game is shown in Figure \ref{fig_scheme_game} in Appendix \ref{app:pre}.


\paragraph{Learning Approach} We employ \textbf{Reinforcement Learning from Ranking Feedback (RLRF)} to train our agent, henceforth referred to as {\bf RL-aligned agent (RA agent)}. Specifically, the LLM is trained with signals derived from rankings, enabling it to perform more effectively in ranking competitions at test time. To this end, we generate synthetic data to construct a preference dataset\footnote{A preference dataset consists of triplets: (i) a prompt or feedback context, (ii) a positive example (a document modification that is ranked above another candidate), and (iii) a negative example (the lower-ranked candidate); positive/negative labels are derived from the ranker's ordering.} and train the agent to increase the likelihood of content modifications that lead to higher ranks while decreasing the likelihood of those that result in lower ranks. The algorithms implementing RLRF using DPO\footnote{The choice of the DPO algorithm over alternative methods is discussed in Section \ref{sec:agent-train}.} \citep{rafailov2024directpreferenceoptimizationlanguage} are presented in Figure \ref{fig:rlrf_designs}. The key difference between the two algorithms lies in how the documents are generated. In the \textbf{Static Generation (SG; Algorithm \ref{algorithm:shortSG})} setting, for each query, an LLM first generates a pseudo-relevant document to the query, independent of any competitive context. Based on this document, multiple modified variants are then generated using prompts that instruct the LLM to revise the document in different ways\footnote{The prompts are presented in Appendix \ref{app:data-gen-sg}}. A ranking is induced over the resulting pool of documents, and the preference dataset is extracted from the highest- and lowest-ranked variants. This approach enables learning how document modifications influence rankings. In contrast, in the \textbf{Dynamic Generation (DG; Algorithm \ref{algorithm:shortDG})} setting, there is a repeated ranking game, where multiple instances of the same LLM iteratively modify their documents in response to rankings. In this setup, the data generation procedure produces a preference dataset that reflects the evolving competitive dynamics across rounds. Consequently, during training, the algorithm aligns the agent not only with the ranker’s preferences but also with the document-modification strategies that emerge over time in the competition. This alignment enables the agent to adapt its document-modification strategy across rounds and achieve improved performance, as we show in Section \ref{sec:results}. Importantly, our core novelty lies in training an agent for an unknown ranker using only implicit signals induced from rankings, while simultaneously accounting for the strategic behavior of other agents in the competition. Additional algorithmic details are provided in Appendix \ref{appendix:hyperparams}. Preliminaries on RL and DPO are provided in Appendix \ref{app:pre}.

\begin{figure}[t]
\centering
\begin{minipage}[t]{0.49\textwidth}
\begin{algorithm}[H]
\small
\caption{RLRF Agent: Static Generation}
\label{algorithm:shortSG}
\input{algorithms/shortSG2}
\end{algorithm}
\end{minipage}\hfill
\begin{minipage}[t]{0.49\textwidth}
\begin{algorithm}[H]
\small
\caption{RLRF Agent: Dynamic Generation}
\label{algorithm:shortDG}
\input{algorithms/shortDG2}
\end{algorithm}
\end{minipage}
\caption{RLRF Agent Designs: Static (left) vs. Dynamic (right).}
\label{fig:rlrf_designs}
\end{figure}

%
%
\section{Experimental Setting}\label{sec:exp}

In this section, we detail the framework employed for training and evaluating the \rl{}. The agent is trained on the synthetic preference datasets derived from a simulated ranking competition between large language models (LLMs). To train our agent, we adopt Direct Preference Optimization (DPO; \citealp{rafailov2024directpreferenceoptimizationlanguage}), utilizing a set of prompts introduced in prior work \citep{bardas_automatic_2025}. The performance of the resulting agent is evaluated using the \testEnv{} simulated environment for LLM-based ranking competitions \citep{mordo_lemss_2025}.


\subsection{Components}\label{sec:components}
\paragraph{LLMs and Prompts} 
We used lightweight instruct-tuned language models (< 10B parameters) as our agents: Llama3.1 \citep{dubey_llama_2024}, Mistral \citep{jiang_mistral_2023}, Gemma2 \citep{gemma_team_gemma_nodate}, and Qwen2.5\footnote{Models sourced from the Hugging Face repository: meta-llama/Meta-Llama-3.1-8B-Instruct, mistralai/Mistral-8B-Instruct-2410, google/gemma-2-9b, and Qwen/Qwen2.5-7B-Instruct.} \citep{qwen_qwen25_2024}. The choice of LLMs was motivated by two reasons. First, using lightweight models allows us to conduct large-scale training and evaluation under reasonable computational constraints \citep{belcak_small_2025}. Second, this setup aligns, and therefore allows comparison, with prior work on competitive search \citep{mordo_lemss_2025}, where models with up to 10 billion parameters were used to ensure reproducibility and accessibility \citep{belcak_small_2025}. The prompts used in our experiments are from \cite{bardas_automatic_2025}; LLM-based agents guided by these prompts consistently outperformed student participants in single-round document modification. Specifically, we employ (i) the Pairwise Prompt agent (\textbf{\nivPair{}}) and (ii) the Listwise Prompt agent (\textbf{\nivList{}}) \citep{bardas_automatic_2025}. The \nivPair{} prompt consists of the last three rounds of a pair of documents and their ranks with respect to the query. The \nivList{} prompt consists of the last two rounds of the entire ranked list with respect to the query. We denote these prompt-based agents as {\bf non-aligned agents (NA agents)}, since they were not trained prior to the ranking competition but rather calibrated only through hyper-parameter tuning and prompt engineering.

\paragraph{Ranking Functions}
We employed three dense retrieval ranking functions and one sparse retrieval method. The dense rankers, following prior work on ranking competitions \citep{mordo_lemss_2025}, are: E5 in both its unsupervised and supervised variants \citep{wang_text_2024}, and Contriever\footnote{The dense models were obtained from the Hugging Face repository: intfloat/e5-large-unsupervised, intfloat/e5-large-supervised, and facebook/contriever.} \citep{izacard_unsupervised_2022}. The sparse ranker is Okapi BM25 \citep{robertson1993okapi}. For the dense retrieval models, ranking scores for document–query pairs were computed using cosine similarity between their respective embedding vectors. For the BM25 ranking function, we extracted inverse document frequency (IDF) features from a 59,000-document subset of the English Wikipedia, based on a 2020 dump. The text was normalized using Krovetz stemming, following the pre-processing protocol described in \cite{Frej2020MlWikir,Frej2020Wikir}.


\paragraph{Queries and Initial Documents}
Each game is assigned with a query for which the agents compete. The game begins with the same initial document that each agent is required to modify in an effort to improve its ranking for the given query. We selected 500 queries from the Passage Ranking task of the TREC 2022 test collection, which is based on the MS MARCO dataset \citep{Bajaj2016Msmarco,craswell_overview_2025}; the queries were divided randomly to 90\% for the training dataset and 10\% for the test dataset. For each query, we also selected an initial document from the MS MARCO Passage collection that had been manually judged as highly relevant to that query\footnote{Three out of three annotators judged the document as relevant to the query.}. The documents are therefore short as in prior studies of competitive search  \citep{raifer2017information}.

\subsection{Data Generation}\label{sec:data}

Recent work demonstrated remarkable success in improving the performance of AI models using synthetic data in strategic decision-making \citep{shapira2024can,shapira2025human} and gaming scenarios \citep{silver_mastering_2017, silver_general_2018}. Inspired by this line of research,
we constructed synthetic datasets to train and optimize LLM-based agents in our competitive search setting. In alignment with real-world scenarios, where Web publishers typically do not have knowledge of the internals of ranking algorithms, we assume that agents are exposed only to the ranked list of documents. The use of generative AI to construct preference datasets tailored to task-specific fine-tuning of language models has been studied in prior work \citep{bai2022constitutionalaiharmlessnessai, lee2024rlaifvsrlhfscaling,gao2024unifiedviewpreferencelearning}. Inspired by this line of research, we generate training data by sampling outputs from a ranker using two methods: Static Generation (\staticDocsGeneration{}) and Dynamic Generation (\dynamicDocsGeneration{}) as discussed in Section \ref{sec:task}. More technical details are provided in Appendices \ref{app:data-gen} and \ref{appendix:gen-hyper}.

\subsection{Agent Training}\label{sec:agent-train}


We train the \rl{}s using the data generation methods introduced in Section \ref{sec:data}. In line with prior work on competitive search, we instruct the agents to generate short documents of approximately 150 words \citep{bardas_automatic_2025,mordo_lemss_2025}. In contrast to RLHF \citep{christiano2017deep}, which aligns model outputs with human preferences, our objective is to align agent behavior (namely, document modification strategies) with the preferences of a ranker. Importantly, the agent is only exposed to rankings for a limited set of queries, without access to scores or model internals.
Rather than relying on less stable optimization methods such as Proximal Policy Optimization (PPO; \citealp{schulman2017proximalpolicyoptimizationalgorithms}), which typically require training an explicit reward model and collecting a large dataset to approximate the behavior of a ranker, we adopt Direct Preference Optimization\footnote{Exploring alternative optimization methods, such as GRPO \citep{guo_deepseek-r1_2025}, is left for future work.} (DPO; \citealp{rafailov2024directpreferenceoptimizationlanguage}). DPO offers a more stable and sample-efficient alternative, as it directly optimizes model parameters using pairwise preference data \citep{ wu2023pairwiseproximalpolicyoptimization,rafailov2024directpreferenceoptimizationlanguage}. Each training example consists of a prompt, a preferred (positive) document, and a less-preferred (negative) document. The loss encourages the model to assign higher likelihood to preferred documents\footnote{In our setup, the positive document corresponds to the top-ranked output, while the negative document is the lowest-ranked one, as determined by the ranker.}. This formulation allows for effective alignment with ranking-based preferences without explicitly modeling the reward function. A detailed description of the training setup and hyper-parameters is provided in Appendix \ref{appendix:train}.

\subsection{Evaluation}\label{sec:eval}





Our setting models repeated interactions where agents iteratively modify their documents over multiple rounds in response to ranking and the strategic behavior of other agents. We present two evaluation settings: {\bf Homogeneous (denoted \homoEval{}) and Heterogeneous (denoted \hetroEval{})}. In the \homoEval{} setting the \rl{} competes against duplications of \prompt{}s (non-aligned agents) with the same language models as the \rl{}. In the \hetroEval{} setting the \rl{} competes against \prompt{}s with different language models. Recall that the feedback to all the agents is provided by using the \nivList{} or the \nivPair{} prompts \citep{bardas_automatic_2025}. For each setting, we compare the \textbf{win-rate}\footnote{A win means being ranked the highest for a round.} of the \rl{} against the \textit{best} performing \prompt{} for that specific setting\footnote{A subtle consideration arises in the \homoEval{} setup. Since the opponents are identical \prompt{}s, their wins are distributed equally among them. This can lead to an extreme case in which the \rl{} performs exactly the same as every instance of the \prompt{}, yet — because of the duplication of opponents — it appears that the \rl{} outperforms each of them individually. To account for this effect, we include in Appendix \ref{ape:robust} a dedicated 1-vs-1 competition between the \rl{} and a \prompt{}.}. We evaluate an agent performance in the ranking competition simulated using \testEnv{} \citep{mordo_lemss_2025} measuring the win-rate averaged across games in the competition. We also define a \textit{random baseline} whose performance is the expected win-rate if all agents have an equal probability of winning each round (i.e., $1/k$ for $k$ competing agents). See Appendix \ref{appendix:eval} for detailed description of the measures. Statistical significance is measured using a two-tailed paired permutation test with $p = 0.05$ and $10,000$ permutations.

In addition to win-rate, we evaluate the \textbf{faithfulness} of the modified documents to their original counterparts in order to capture cases of substantial modifications made in pursuit of ranking promotion. Following \citet{bardas_automatic_2025}, we employ an NLI model developed by \citet{gekhman2023trueteacherlearningfactualconsistency} to compute whether a modified document is entailed by the initial document. A formal definition of this measure is provided in Appendix \ref{appendix:eval}.
\section{Analysis and Results}\label{sec:results}

We begin by presenting the research questions (RQs) that guide the evaluation of the \rl{}s. For each RQ, we define one or more experimental settings that enable a comprehensive analysis of the agent's behavior and performance:

\begin{itemize}[noitemsep, topsep=-4pt, leftmargin=*]
\item \textbf{RQ1:} To what extent does the \rl{} outperform \prompt{}s in repeated ranking competitions between LLMs?
\item \textbf{RQ2:} How well does the \rl{} generalize to unseen ranking functions, and how robust is it to potential misalignment between training and test-time ranking functions?

\end{itemize}

We evaluate the \rl{} (compared to \prompt{}) in simulated ranking competitions. For RQ1 we use the two configurations \homoEval{} and \hetroEval{}. For RQ2, we used the \hetroEval{} setup, as it is considered more challenging for the \rl{}. This setup incorporates the \rl{} alongside the multiple \prompt{}s with different language models. We used the \rl{} built on Mistral, trained with \dynamicDocsGeneration{} and prompted with \nivList{}, since it achieved the highest win-rate in RQ1. This choice was driven by the limited resources available for training, which required us to focus subsequent experiments on one agent configuration.
Each competition consists of 50 games, initialized with a query not used in the training set and a corresponding initial document. Each game spans 30 rounds, which prior work has shown to be sufficient for convergence in LLM-based ranking competitions \citep{mordo_lemss_2025}.




\subsection{RQ1: effectiveness of the \rl{} in ranking competition}\label{sec:rq1}
To address RQ1, we evaluate the effectiveness of our \rl{} in comparison to \prompt{}s in a ranking competition that is conducted over multiple rounds. The evaluation is conducted in the \testEnv{} simulator for ranking competitions. We trained four lightweight language models: Mistral, Gemma, Llama, and Qwen. We used two distinct data generation methods: \staticDocsGeneration{} (Static Generation) and \dynamicDocsGeneration{} (Dynamic Generation); see Section \ref{sec:task} for more details on these generation methods. In \staticDocsGeneration{}, the pseudo-relevant document for each query was modified five times. In the \dynamicDocsGeneration{} setup, we first simulated a competition with 450 games (one game per query), each consisting of 30 rounds and five instantiations of \prompt{}s. We used the generated documents as a training dataset. For both generation methods we used a temperature of 0.8 \citep{yuan_rrhf_2023}.
Consistent with prior work \citep{mordo2025searchresultsdiversificationcompetitive, bardas_automatic_2025}, we employed \nivPair{} and \nivList{} as the prompting strategies, and used the unsupervised E5 ranking function \citep{wang_text_2024} for both data generation and evaluation.

Gemma, Llama, and Qwen were trained only under the \dynamicDocsGeneration{} setup with the \nivList{} prompt following initial evaluation in which we ran a competition with the base (non-RL) versions of all four models. Mistral was the worst-performing model in this initial evaluation, and was therefore selected for a broader configuration analysis, including the \staticDocsGeneration{} and the \nivPair{} prompt. The motivation for focusing on Mistral was to demonstrate that even if the underlying LLM performs the worst in an initial evaluation, it is still possible to design an \rl{} that outperforms \prompt{}s based on other LLMs.


Table \ref{tab:rq2} presents the win-rate comparisons across different competition configurations. In all cases, the \rl{} outperforms the random baseline (20\% win-rate). Moreover, in nearly all scenarios, the \rl{} significantly outperformed the best \prompt{}\footnote{Except for the case of Llama trained with \dynamicDocsGeneration{}, using the \nivList{} prompt and evaluated under the \hetroEval{} setting}. Notably, the \rl{} fine-tuned on Mistral with the \nivList{} prompt achieved the highest win rates under both \homoEval{} and \hetroEval{} settings ($0.75$ and $0.6$, respectively).
Among agents trained with \dynamicDocsGeneration{}, Table \ref{tab:rq2} shows consistently higher performance in the \homoEval{} setting compared to \hetroEval{}. This can be attributed to the alignment between the agent's underlying language model and those used by its competitors and for data generation in the \homoEval{} case. In contrast, the \hetroEval{} setting includes heterogeneous agents based on different underlying LLMs, thereby introducing more diverse documents that challenge our \rl{} to adapt its strategy effectively. In Appendix \ref{ape:robust}, we extend our analysis and demonstrate that the performance of the \rl{} remains robust with respect to both the number of competitors and the evaluation-time temperature of the LLM.

A comparison between \staticDocsGeneration{} and \dynamicDocsGeneration{} in Table \ref{tab:rq2} highlights two distinct aspects of the designing of the \rl{}. \staticDocsGeneration{} primarily focuses on aligning the agent with the ranker by learning which document variants are preferred, but it does not account for the evolving strategies of other competitors. In contrast, \dynamicDocsGeneration{} explicitly models the dynamic nature of the task by simulating multi-round competitions in which agents continuously adapt their modifications in response to rankings. This distinction is reflected in Table \ref{tab:rq2}, where \dynamicDocsGeneration{}-trained agents consistently outperform their \staticDocsGeneration{} counterparts -- most notably for Mistral. The \rl{} in the setting with the \nivList{} prompt and \dynamicDocsGeneration{} procedure achieves a win-rate of $0.60$ under \hetroEval{} setting and $0.75$ under the \homoEval{} setting, compared to $0.29$ for \staticDocsGeneration{} in \hetroEval{}. For the \nivPair{} prompt the trends are similar. \staticDocsGeneration{} under the \hetroEval{} setting achieves a win-rate of $0.29$ while the \dynamicDocsGeneration{} achieves $0.36$. These results indicate that designing a competitive agent cannot be reduced to the static task of learning the ranker alone; rather, it also requires learning effective strategies against adaptive opponents.

To further contextualize these findings, we additionally explored the document-modification strategies employed by the agents in Appendix \ref{app:strategies}. Our analysis revealed that in the \dynamicDocsGeneration{} setting, greater diversity in ranked lists was observed for the \rl{} compared to the \prompt{}. This effect arises because the \rl{} makes more substantial document modifications across rounds, leading also to lower similarity between successive documents' versions than in the \prompt{}s. In contrast, the \staticDocsGeneration{} setting yields more homogeneous documents and similar modification patterns for both \rl{} and \prompt{}. Consistent with prior work \cite{mordo_lemss_2025}, both agent types eventually converge toward stable documents.

We also analyzed how competition affects both the win-rate and the relevance judgments of the \rl{} and \prompt{}s (Appendices \ref{app:strategies} and \ref{app:annotation}). Relevance annotations and win-rate analyses show that the stronger alignment of the \rl{} with the ranker provides a clear advantage at the start of the competition: in round 1, the \rl{} produces documents of significantly higher relevance and achieves higher win-rates than the \prompt{}. By round 30, however, this advantage reduced as \prompt{}s improve through competition — an instance of the herding effect \cite{raifer2017information}, where all agents gravitate toward similar highly relevant documents. Notably, in the \staticDocsGeneration{} setting, the advantage of the \rl{} relative to the \prompt{} is substantially reduced compared to \dynamicDocsGeneration{}.

We extend the study to competitions involving multiple \rl{}s in Appendix \ref{app:multi-rlrf}. When multiple \rl{}s compete, their presence increases the inter-document similarity in ranked lists but does not significantly affect overall ranking performance, suggesting that diverse adaptation strategies primarily emerge in multiple-agents settings.

Finally, in Appendix \ref{app:rq3new}, we further evaluate our agent in the single-round setting of \citet{bardas_automatic_2025}. The results show that our \rl{} consistently outperforms the \prompt{}s in both ranking promotion and content faithfulness. Together with the repeated-competition evaluation, these findings demonstrate that the advantages of our \rl{} extend across several competitive settings.

\begin{table}[t]
\caption{Comparison of agent performance under heterogeneous (\hetroEval{}) and homogeneous (\homoEval{}) configurations. We report the win-rate (\textbf{WR}) of the \rl{} (RL-aligned agent) and the best \prompt{} (non-aligned agent). '$*$' marks a statistically significant difference with the win-rate of the best \prompt{} in the same configuration. The best performance in each configuration is boldfaced.}
\centering
\small

\begin{tabular}{ll|cc|cc}
\toprule
\textbf{LLM} & \textbf{Train Setting} 
& \multicolumn{2}{c|}{\textbf{Heterogeneous}} 
& \multicolumn{2}{c}{\textbf{Homogeneous}} \\
\midrule
 &  &  \makecell[c]{\textbf{\rl{}}\\\textbf{WR}}  & \makecell[c]{\textbf{Best \prompt{}}\\\textbf{WR}}  & \makecell[c]{\textbf{\rl{}}\\\textbf{WR}}  & \makecell[c]{\textbf{Best \prompt{}}\\\textbf{WR}} \\
\midrule
Mistral & \staticDocsGeneration{} (\nivPair{})     & $\mathbf{0.29^{*}}$ & $0.21$ & $\mathbf{0.29^{*}}$ & $0.2$     \\
Mistral & \staticDocsGeneration{} (\nivList{})     & $\mathbf{0.29^{*}}$ & $0.20$ & $\mathbf{0.58^{*}}$ & $0.17$     \\
Mistral & \dynamicDocsGeneration{} (\nivPair{})    & $\mathbf{0.36^{*}}$ & $0.20$ & $\mathbf{0.71^{*}}$ & $0.13$     \\
Mistral & \dynamicDocsGeneration{} (\nivList{})    & $\mathbf{0.60^{*}}$ & $0.11$ & $\mathbf{0.75^{*}}$ & $0.10$ \\
Gemma   & \dynamicDocsGeneration{} (\nivList{})    & $\mathbf{0.34^{*}}$ & $0.19$ & $\mathbf{0.54^{*}}$ & $0.15$ \\
Llama   & \dynamicDocsGeneration{} (\nivList{})    & $0.24$     & $0.24$ & $\mathbf{0.59^{*}}$ & $0.14$ \\
Qwen    & \dynamicDocsGeneration{} (\nivList{})    & $\mathbf{0.33^{*}}$ & $0.18$ & $\mathbf{0.49^{*}}$ & $0.16$ \\
\bottomrule
\end{tabular}
\label{tab:rq2}
\end{table}

\begin{figure}[t]
\centering
\small
\includegraphics[width=0.90\linewidth]{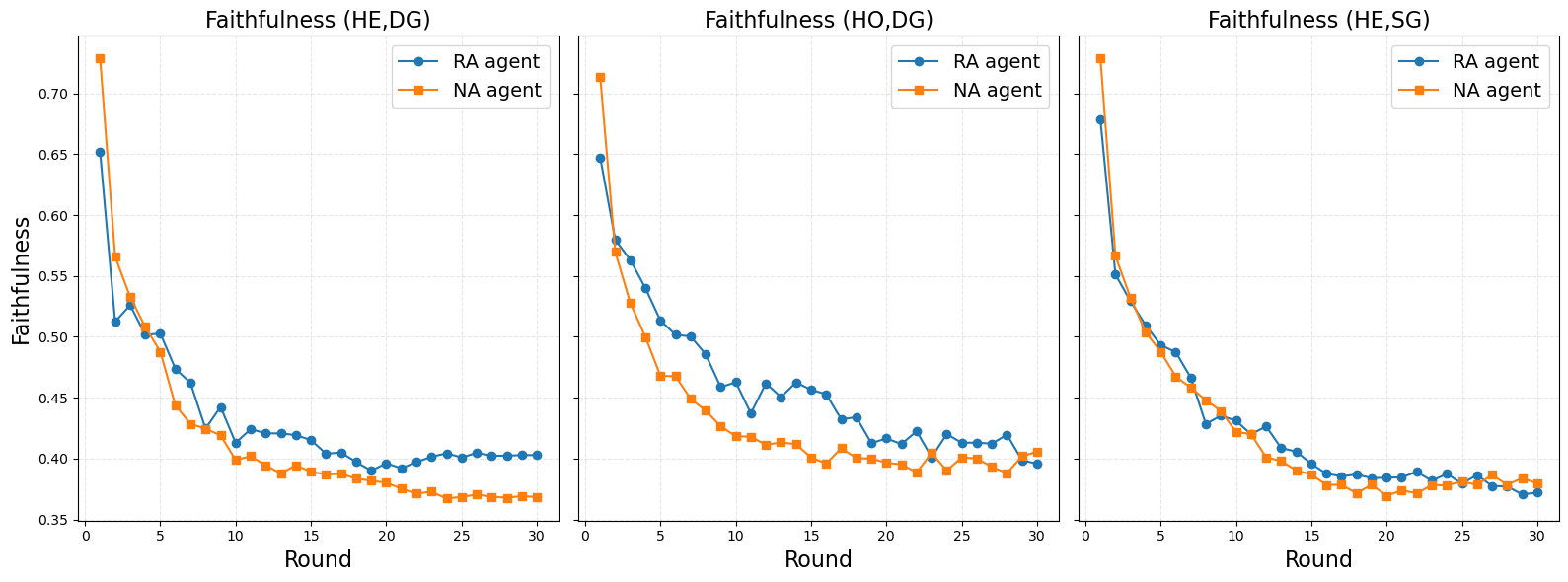}
\caption{The faithfulness score of the \rl{} and the \prompt{} for the \hetroEval{} and \dynamicDocsGeneration{} (left), \homoEval{} and \dynamicDocsGeneration{} (middle), and \hetroEval{} and \staticDocsGeneration{} (right) settings.}
\label{fig:faith-rq2}
\end{figure}

\paragraph{Faithfulness Analysis} We analyzed the faithfulness of agents-modified documents to their original versions over 30 competition rounds, averaging scores across queries. We evaluated the \rl{} and the \prompt{} in the configurations prompted with \nivList{} and instantiated with the Mistral language model. The configurations included \dynamicDocsGeneration{} under both \hetroEval{} and \homoEval{}, and \staticDocsGeneration{} under \hetroEval{}. The comparison of the faithfulness between the \rl{} and the \prompt{} is shown in Figure \ref{fig:faith-rq2}. In the early rounds (Rounds 1–4), both agents in all settings maintain relatively high faithfulness, with scores above 0.5\footnote{I.e., more than 50\% of the sentences are entailed by the initial document \citep{gekhman2023trueteacherlearningfactualconsistency}.}. In addition, across most rounds, the \rl{} consistently achieves higher faithfulness than the \prompt{}. Toward later rounds, both agents exhibit converging faithfulness trends, reflecting limited further document modifications, a phenomenon consistent with observations in prior work \citep{mordo_lemss_2025}. 

Overall, our results suggest that the \rl{} not only outperforms the \prompt{} in win-rate, but also better preserves the faithfulness to the original document throughout the competition.



\subsection{RQ2: transfer learning across ranking functions}

In RQ2, we study the extent to which the performance of the \rl{} generalizes across ranking functions, specifically when there is a mismatch between the ranker used during training and the one used during evaluation. This setting reflects realistic deployment scenarios, where the true ranking function may differ from the one used during development or may even change over time. Hence, robustness to ranker shifts is a key requirement for practical applicability. We focus on the best \rl{} from RQ1: the Mistral language model, trained using the \dynamicDocsGeneration{} procedure and the \nivList{} prompt.

We trained the agent using two different ranking functions: E5-unsupervised \citep{wang_text_2024}, and Contriever \citep{izacard_unsupervised_2022}. Evaluation was conducted under the \hetroEval{} competition setting, using each of the aforementioned rankers as well as two additional rankers: (1) a supervised variant of E5 \citep{wang_text_2024}, to study the impact of supervision in the ranking function, and (2) Okapi BM25 \citep{robertson1993okapi}.

Table \ref{tab:rq3} presents the win-rate results of the \rl{} and the \prompt{} across the various combinations of training and evaluation ranking functions. In almost all relevant comparisons, the \rl{} significantly outperformed the best \prompt{} in the competition, attesting to its ability to transfer effectively across rankers, even when they were not used for training. Interestingly, the results reveal that transfer learning across ranking functions is asymmetric. For instance, when the \rl{} is trained using the E5-unsupervised ranker and evaluated on Contriever, it achieves a win-rate of $0.28$. In contrast, when trained with Contriever and evaluated using E5-unsupervised, the win-rate increases to $0.50$. This asymmetry suggests that certain rankers may induce more generalizable training signals than others.
All in all, these findings highlight both the robustness and the directional sensitivity of transfer learning of our \rl{}s in repeated ranking games. 


\begin{table}[t]
\caption{Comparison of the win-rate (\textbf{WR}) in the \hetroEval{} competitions with Mistral 8B agents trained with \dynamicDocsGeneration{} and prompted with \nivList{} under different ranking functions used for training and evaluation. We report the win-rate of the \rl{} (RL-aligned agent) and the best \prompt{} (non-aligned agent). '*' marks a statistically significant difference with the win-rate of the best \prompt{}.}
\centering
\small

\begin{tabular}{lllc|cc}
\toprule
\textbf{LLM} & \textbf{Train Setting} & \textbf{Trained Ranker} & \textbf{Tested Ranker} 
& \makecell[c]{\textbf{\rl{}}\\\textbf{WR}} & \makecell[c]{\textbf{Best \prompt{}}\\\textbf{WR}} \\
\midrule
\multirow{3}{*}{Mistral} & 
\multirow{6}{*}{\dynamicDocsGeneration{} (\nivList{})} & 
\multirow{3}{*}{E5-unsupervised} & 
E5-supervised & $\mathbf{0.27^{*}}$ & $0.20$ \\
&&& Contriever & $\mathbf{0.28}$ & $0.25$ \\
&&& Okapi & $\mathbf{0.29^{*}}$ & $0.21$ \\
\cmidrule(l){3-6}
&& \multirow{3}{*}{Contriever} & E5-supervised & $\mathbf{0.44^{*}}$ & $0.17$ \\
&&& E5-unsupervised & $\mathbf{0.50^{*}}$ & $0.15$ \\
&&& Okapi & $\mathbf{0.58^{*}}$ & $0.12$ \\
\bottomrule
\end{tabular}

\label{tab:rq3}
\end{table}


\section{Conclusion}
We introduced an RL-aligned (RA) agent for competitive search, where LLMs act as publishers in repeated ranking games. Our extensive experiments show that our agent consistently outperforms non-aligned (NA) agents, demonstrating the effectiveness of RL in this strategic retrieval setting.
For future work, we intend to pursue several directions. First, devising alternative optimization strategies and loss formulations specifically tailored to ranking-based alignment is a promising avenue for improving agent performance. Second, we plan to design RL-based strategies that explicitly encourage higher levels of faithfulness, with the goal of balancing ranking effectiveness and faithfulness to the original document. Finally, we aim to explore online agents that can learn and adapt during the ranking competition itself, rather than being trained solely before test time.

\paragraph{Ethics Statement}
This research does not involve human subjects, personal data, or sensitive information, and therefore does not raise privacy, security, or IRB-related concerns. All datasets used are publicly available (e.g., MS MARCO, TREC) or synthetically generated by large language models, and no copyrighted or proprietary data was included. Our experiments focus on ranking competitions in a controlled simulation framework and do not involve deployment in real-world systems. While our work introduces reinforcement learning strategies to optimize LLM-based agents in competitive search, we acknowledge that ranking manipulation and strategic content generation may raise concerns if misused. To mitigate such risks, we restrict our study to academic evaluation settings and will release in the camera-ready version code and data solely for reproducibility and further research in information retrieval and responsible AI. We also note that both large language models and ranking functions may reflect societal biases present in their training data. Although addressing bias and fairness is not the primary focus of this work, we encourage future studies to examine how such factors interact with strategic content generation in competitive search.

\paragraph{Reproducibility Statement} We provide a detailed description of our algorithms in Section \ref{sec:task}, with additional technical details in Appendices \ref{app:data-gen} and \ref{appendix:train}. Hyper-parameters for dataset generation and agent training are reported in Appendix \ref{appendix:hyperparams}. All evaluation measures are well-defined (see Appendix \ref{appendix:eval}) to facilitate replication. The datasets we used for evaluation, as well as the code for analysis, data generation, and agent design will be released with the camera-ready version to ensure full reproducibility.

\bibliographystyle{plainnat}

\begin{thebibliography}{94}
\providecommand{\natexlab}[1]{#1}
\providecommand{\url}[1]{\texttt{#1}}
\expandafter\ifx\csname urlstyle\endcsname\relax
  \providecommand{\doi}[1]{doi: #1}\else
  \providecommand{\doi}{doi: \begingroup \urlstyle{rm}\Url}\fi

\bibitem[noa(2024)]{noauthor_introducing_2024}
Introducing {Connect} by {CloudResearch}: {Advancing} {Online} {Participant} {Recruitment} in the {Digital} {Age} {\textbar} {Request} {PDF}, July 2024.
\newblock URL \url{https://www.researchgate.net/publication/373983592_Introducing_Connect_by_CloudResearch_Advancing_Online_Participant_Recruitment_in_the_Digital_Age}.

\bibitem[Akata et~al.(2025)Akata, Schulz, Coda-Forno, Oh, Bethge, and Schulz]{akata2025playing}
Elif Akata, Lion Schulz, Julian Coda-Forno, Seong~Joon Oh, Matthias Bethge, and Eric Schulz.
\newblock Playing repeated games with large language models.
\newblock \emph{Nature Human Behaviour}, pages 1--11, 2025.

\bibitem[Alice(2006)]{alice2006manipulability}
CHENG Alice.
\newblock Manipulability of pagerank under sybil strategies.
\newblock In \emph{First Workshop on the Economics of Networked Systems (NetEcon06)}, 2006.

\bibitem[Bai et~al.(2022{\natexlab{a}})Bai, Kadavath, Kundu, Askell, Kernion, Jones, Chen, Goldie, Mirhoseini, McKinnon, Chen, Olsson, Olah, Hernandez, Drain, Ganguli, Li, Tran-Johnson, Perez, Kerr, Mueller, Ladish, Landau, Ndousse, Lukosuite, Lovitt, Sellitto, Elhage, Schiefer, Mercado, DasSarma, Lasenby, Larson, Ringer, Johnston, Kravec, Showk, Fort, Lanham, Telleen-Lawton, Conerly, Henighan, Hume, Bowman, Hatfield-Dodds, Mann, Amodei, Joseph, McCandlish, Brown, and Kaplan]{bai2022constitutionalaiharmlessnessai}
Yuntao Bai, Saurav Kadavath, Sandipan Kundu, Amanda Askell, Jackson Kernion, Andy Jones, Anna Chen, Anna Goldie, Azalia Mirhoseini, Cameron McKinnon, Carol Chen, Catherine Olsson, Christopher Olah, Danny Hernandez, Dawn Drain, Deep Ganguli, Dustin Li, Eli Tran-Johnson, Ethan Perez, Jamie Kerr, Jared Mueller, Jeffrey Ladish, Joshua Landau, Kamal Ndousse, Kamile Lukosuite, Liane Lovitt, Michael Sellitto, Nelson Elhage, Nicholas Schiefer, Noemi Mercado, Nova DasSarma, Robert Lasenby, Robin Larson, Sam Ringer, Scott Johnston, Shauna Kravec, Sheer~El Showk, Stanislav Fort, Tamera Lanham, Timothy Telleen-Lawton, Tom Conerly, Tom Henighan, Tristan Hume, Samuel~R. Bowman, Zac Hatfield-Dodds, Ben Mann, Dario Amodei, Nicholas Joseph, Sam McCandlish, Tom Brown, and Jared Kaplan.
\newblock Constitutional ai: Harmlessness from ai feedback, 2022{\natexlab{a}}.
\newblock URL \url{https://arxiv.org/abs/2212.08073}.

\bibitem[Bai et~al.(2022{\natexlab{b}})Bai, Kadavath, Kundu, Askell, Kernion, Jones, Chen, Goldie, Mirhoseini, McKinnon, et~al.]{bai2022constitutional}
Yuntao Bai, Saurav Kadavath, Sandipan Kundu, Amanda Askell, Jackson Kernion, Andy Jones, Anna Chen, Anna Goldie, Azalia Mirhoseini, Cameron McKinnon, et~al.
\newblock Constitutional ai: Harmlessness from ai feedback.
\newblock \emph{arXiv preprint arXiv:2212.08073}, 2022{\natexlab{b}}.

\bibitem[Bar-Ilan(2007)]{bar2007manipulating}
Judit Bar-Ilan.
\newblock Manipulating search engine algorithms: the case of google.
\newblock \emph{Journal of Information, Communication and Ethics in Society}, 5\penalty0 (2/3):\penalty0 155--166, 2007.

\bibitem[Bardas et~al.(2025)Bardas, Mordo, Kurland, and Tennenholtz]{bardas_automatic_2025}
Niv Bardas, Tommy Mordo, Oren Kurland, and Moshe Tennenholtz.
\newblock Automatic {Document} {Editing} for {Improved} {Ranking}.
\newblock In \emph{Proceedings of the 48th {International} {ACM} {SIGIR} {Conference} on {Research} and {Development} in {Information} {Retrieval}}, {SIGIR} '25, pages 2779--2783, New York, NY, USA, July 2025. Association for Computing Machinery.
\newblock ISBN 9798400715921.
\newblock \doi{10.1145/3726302.3730168}.
\newblock URL \url{https://dl.acm.org/doi/10.1145/3726302.3730168}.

\bibitem[Beirami et~al.(2025)Beirami, Agarwal, Berant, D'Amour, Eisenstein, Nagpal, and Suresh]{beirami2025theoreticalguaranteesbestofnalignment}
Ahmad Beirami, Alekh Agarwal, Jonathan Berant, Alexander D'Amour, Jacob Eisenstein, Chirag Nagpal, and Ananda~Theertha Suresh.
\newblock Theoretical guarantees on the best-of-n alignment policy, 2025.
\newblock URL \url{https://arxiv.org/abs/2401.01879}.

\bibitem[Belcak et~al.(2025)Belcak, Heinrich, Diao, Fu, Dong, Muralidharan, Lin, and Molchanov]{belcak_small_2025}
Peter Belcak, Greg Heinrich, Shizhe Diao, Yonggan Fu, Xin Dong, Saurav Muralidharan, Yingyan~Celine Lin, and Pavlo Molchanov.
\newblock Small {Language} {Models} are the {Future} of {Agentic} {AI}, June 2025.
\newblock URL \url{http://arxiv.org/abs/2506.02153}.
\newblock arXiv:2506.02153 [cs].

\bibitem[Ben~Basat et~al.(2015)Ben~Basat, Tennenholtz, and Kurland]{ben2015probability}
Ran Ben~Basat, Moshe Tennenholtz, and Oren Kurland.
\newblock The probability ranking principle is not optimal in adversarial retrieval settings.
\newblock In \emph{Proceedings of the 2015 International Conference on The Theory of Information Retrieval}, pages 51--60, 2015.

\bibitem[Ben~Basat et~al.(2017)Ben~Basat, Tennenholtz, and Kurland]{basat2017game}
Ran Ben~Basat, Moshe Tennenholtz, and Oren Kurland.
\newblock A game theoretic analysis of the adversarial retrieval setting.
\newblock \emph{Journal of Artificial Intelligence Research}, 60:\penalty0 1127--1164, 2017.

\bibitem[Ben-Porat et~al.(2019)Ben-Porat, Rosenberg, and Tennenholtz]{ben2019convergence}
Omer Ben-Porat, Itay Rosenberg, and Moshe Tennenholtz.
\newblock Convergence of learning dynamics in information retrieval games.
\newblock In \emph{Proceedings of the AAAI Conference on Artificial Intelligence}, volume~33, pages 1780--1787, 2019.

\bibitem[Brown et~al.(2020)Brown, Mann, Ryder, Subbiah, Kaplan, Dhariwal, Neelakantan, Shyam, Sastry, Askell, et~al.]{brown2020language_lang1}
Tom Brown, Benjamin Mann, Nick Ryder, Melanie Subbiah, Jared~D Kaplan, Prafulla Dhariwal, Arvind Neelakantan, Pranav Shyam, Girish Sastry, Amanda Askell, et~al.
\newblock Language models are few-shot learners.
\newblock \emph{Advances in neural information processing systems}, 33:\penalty0 1877--1901, 2020.

\bibitem[Christiano et~al.(2017)Christiano, Leike, Brown, Martic, Legg, and Amodei]{christiano2017deep}
Paul~F Christiano, Jan Leike, Tom Brown, Miljan Martic, Shane Legg, and Dario Amodei.
\newblock Deep reinforcement learning from human preferences.
\newblock \emph{Advances in neural information processing systems}, 30, 2017.

\bibitem[Coppolillo et~al.(2024)Coppolillo, Cinus, Minici, Bonchi, and Manco]{coppolillo2024engagement}
Erica Coppolillo, Federico Cinus, Marco Minici, Francesco Bonchi, and Giuseppe Manco.
\newblock Engagement-driven content generation with large language models.
\newblock \emph{arXiv preprint arXiv:2411.13187}, 2024.

\bibitem[Craswell et~al.(2025)Craswell, Mitra, Yilmaz, Campos, Lin, Voorhees, and Soboroff]{craswell_overview_2025}
Nick Craswell, Bhaskar Mitra, Emine Yilmaz, Daniel Campos, Jimmy Lin, Ellen~M. Voorhees, and Ian Soboroff.
\newblock Overview of the {TREC} 2022 deep learning track, July 2025.
\newblock URL \url{http://arxiv.org/abs/2507.10865}.
\newblock arXiv:2507.10865 [cs].

\bibitem[Drivas et~al.(2017)Drivas, Sarlis, Sakas, and Varveris]{drivas2017stuffing}
Ioannis~C Drivas, Apostolos~S Sarlis, Damianos~P Sakas, and Alexandros Varveris.
\newblock Stuffing keyword regulation in search engine optimization for scientific marketing conferences.
\newblock In \emph{Strategic Innovative Marketing: 5th IC-SIM, Athens, Greece 2016}, pages 117--123. Springer, 2017.

\bibitem[Dubey et~al.(2024)Dubey, Jauhri, Pandey, Kadian, Al-Dahle, et~al.]{dubey_llama_2024}
Abhimanyu Dubey, Abhinav Jauhri, Abhinav Pandey, Abhishek Kadian, Al-Dahle, et~al.
\newblock The {Llama} 3 {Herd} of {Models}, August 2024.
\newblock URL \url{http://arxiv.org/abs/2407.21783}.
\newblock arXiv:2407.21783 [cs].

\bibitem[Fleiss(1971)]{fleiss_measuring_1971}
Joseph~L. Fleiss.
\newblock Measuring nominal scale agreement among many raters.
\newblock \emph{Psychological Bulletin}, 76\penalty0 (5):\penalty0 378--382, November 1971.
\newblock ISSN 1939-1455, 0033-2909.
\newblock \doi{10.1037/h0031619}.
\newblock URL \url{https://doi.apa.org/doi/10.1037/h0031619}.

\bibitem[Frej et~al.(2020{\natexlab{a}})Frej, Schwab, and Chevallet]{Frej2020MlWikir}
Jibril Frej, Didier Schwab, and Jean-Pierre Chevallet.
\newblock Mlwikir: A python toolkit for building large-scale wikipedia-based information retrieval datasets in chinese, english, french, italian, japanese, spanish and more.
\newblock In \emph{CIRCLE}, 2020{\natexlab{a}}.

\bibitem[Frej et~al.(2020{\natexlab{b}})Frej, Schwab, and Chevallet]{Frej2020Wikir}
Jibril Frej, Didier Schwab, and Jean-Pierre Chevallet.
\newblock Wikir: A python toolkit for building a large-scale wikipedia-based english information retrieval dataset.
\newblock In \emph{LREC}, 2020{\natexlab{b}}.

\bibitem[Fu et~al.(2023)Fu, Peng, Khot, and Lapata]{fu2023improving}
Yao Fu, Hao Peng, Tushar Khot, and Mirella Lapata.
\newblock Improving language model negotiation with self-play and in-context learning from ai feedback.
\newblock \emph{arXiv preprint arXiv:2305.10142}, 2023.

\bibitem[Gao et~al.(2024{\natexlab{a}})Gao, Song, Miao, Cai, Yang, Chen, Hu, Xu, Dong, Zheng, Quan, Xiao, Zhang, Zan, Lu, Yu, Liu, Cui, Yang, Sha, Wang, Sui, Wang, Liu, and Chang]{gao2024unifiedviewpreferencelearning}
Bofei Gao, Feifan Song, Yibo Miao, Zefan Cai, Zhe Yang, Liang Chen, Helan Hu, Runxin Xu, Qingxiu Dong, Ce~Zheng, Shanghaoran Quan, Wen Xiao, Ge~Zhang, Daoguang Zan, Keming Lu, Bowen Yu, Dayiheng Liu, Zeyu Cui, Jian Yang, Lei Sha, Houfeng Wang, Zhifang Sui, Peiyi Wang, Tianyu Liu, and Baobao Chang.
\newblock Towards a unified view of preference learning for large language models: A survey, 2024{\natexlab{a}}.
\newblock URL \url{https://arxiv.org/abs/2409.02795}.

\bibitem[Gao et~al.(2024{\natexlab{b}})Gao, Song, Miao, Cai, Yang, Chen, Hu, Xu, Dong, Zheng, Quan, Xiao, Zhang, Zan, Lu, Yu, Liu, Cui, Yang, Sha, Wang, Sui, Wang, Liu, and Chang]{gao_towards_2024}
Bofei Gao, Feifan Song, Yibo Miao, Zefan Cai, Zhe Yang, Liang Chen, Helan Hu, Runxin Xu, Qingxiu Dong, Ce~Zheng, Shanghaoran Quan, Wen Xiao, Ge~Zhang, Daoguang Zan, Keming Lu, Bowen Yu, Dayiheng Liu, Zeyu Cui, Jian Yang, Lei Sha, Houfeng Wang, Zhifang Sui, Peiyi Wang, Tianyu Liu, and Baobao Chang.
\newblock Towards a {Unified} {View} of {Preference} {Learning} for {Large} {Language} {Models}: {A} {Survey}, October 2024{\natexlab{b}}.
\newblock URL \url{http://arxiv.org/abs/2409.02795}.
\newblock arXiv:2409.02795 [cs].

\bibitem[Gao et~al.(2024{\natexlab{c}})Gao, Chen, Zhao, Liu, Li, Wang, Zhang, Wang, Ye, Lin, et~al.]{gao2024llm}
Jingtong Gao, Bo~Chen, Xiangyu Zhao, Weiwen Liu, Xiangyang Li, Yichao Wang, Zijian Zhang, Wanyu Wang, Yuyang Ye, Shanru Lin, et~al.
\newblock Llm-enhanced reranking in recommender systems.
\newblock \emph{arXiv preprint arXiv:2406.12433}, 2024{\natexlab{c}}.

\bibitem[Gekhman et~al.(2023)Gekhman, Herzig, Aharoni, Elkind, and Szpektor]{gekhman2023trueteacherlearningfactualconsistency}
Zorik Gekhman, Jonathan Herzig, Roee Aharoni, Chen Elkind, and Idan Szpektor.
\newblock Trueteacher: Learning factual consistency evaluation with large language models, 2023.
\newblock URL \url{https://arxiv.org/abs/2305.11171}.

\bibitem[{Gemma Team} et~al.(){Gemma Team}, {Thomas Mesnard}, {Cassidy Hardin}, {Robert Dadashi}, {Surya Bhupatiraju}, {Laurent Sifre}, {Morgane Rivière}, {Mihir Sanjay Kale}, et~al.]{gemma_team_gemma_nodate}
{Gemma Team}, {Thomas Mesnard}, {Cassidy Hardin}, {Robert Dadashi}, {Surya Bhupatiraju}, {Laurent Sifre}, {Morgane Rivière}, {Mihir Sanjay Kale}, et~al.
\newblock Gemma.
\newblock URL \url{https://www.kaggle.com/m/3301}.

\bibitem[Guo et~al.(2025{\natexlab{a}})Guo, Yang, Zhang, et~al.]{guo_deepseek-r1_2025}
Daya Guo, Dejian Yang, Haowei Zhang, et~al.
\newblock {DeepSeek}-{R1} incentivizes reasoning in {LLMs} through reinforcement learning.
\newblock \emph{Nature}, 645\penalty0 (8081):\penalty0 633--638, September 2025{\natexlab{a}}.
\newblock ISSN 1476-4687.
\newblock \doi{10.1038/s41586-025-09422-z}.
\newblock URL \url{https://doi.org/10.1038/s41586-025-09422-z}.

\bibitem[Guo et~al.(2025{\natexlab{b}})Guo, Li, Zhuang, Luo, Li, Yan, Zhu, and Zhang]{guo2025mcranker}
Fang Guo, Wenyu Li, Honglei Zhuang, Yun Luo, Yafu Li, Le~Yan, Qi~Zhu, and Yue Zhang.
\newblock Mcranker: Generating diverse criteria on-the-fly to improve pointwise llm rankers.
\newblock In \emph{Proceedings of the Eighteenth ACM International Conference on Web Search and Data Mining}, pages 944--953, 2025{\natexlab{b}}.

\bibitem[Guo et~al.(2024{\natexlab{a}})Guo, Bu, Wang, Ren, Sui, Shang, and Lu]{guo2024economics}
Shangmin Guo, Haoran Bu, Haochuan Wang, Yi~Ren, Dianbo Sui, Yuming Shang, and Siting Lu.
\newblock Economics arena for large language models.
\newblock \emph{arXiv preprint arXiv:2401.01735}, 2024{\natexlab{a}}.

\bibitem[Guo et~al.(2024{\natexlab{b}})Guo, Chen, Wang, Chang, Pei, Chawla, Wiest, and Zhang]{guo_large_2024}
Taicheng Guo, Xiuying Chen, Yaqi Wang, Ruidi Chang, Shichao Pei, Nitesh~V. Chawla, Olaf Wiest, and Xiangliang Zhang.
\newblock Large {Language} {Model} {Based} {Multi}-agents: {A} {Survey} of {Progress} and {Challenges}.
\newblock In Kate Larson, editor, \emph{Proceedings of the {Thirty}-{Third} {International} {Joint} {Conference} on {Artificial} {Intelligence}, {IJCAI}-24}, pages 8048--8057. International Joint Conferences on Artificial Intelligence Organization, August 2024{\natexlab{b}}.
\newblock \doi{10.24963/ijcai.2024/890}.
\newblock URL \url{https://doi.org/10.24963/ijcai.2024/890}.

\bibitem[Izacard et~al.(2022)Izacard, Caron, Hosseini, Riedel, Bojanowski, Joulin, and Grave]{izacard_unsupervised_2022}
Gautier Izacard, Mathilde Caron, Lucas Hosseini, Sebastian Riedel, Piotr Bojanowski, Armand Joulin, and Edouard Grave.
\newblock Unsupervised {Dense} {Information} {Retrieval} with {Contrastive} {Learning}, August 2022.
\newblock URL \url{http://arxiv.org/abs/2112.09118}.
\newblock arXiv:2112.09118 [cs].

\bibitem[Jiang et~al.(2023)Jiang, Sablayrolles, Mensch, Bamford, Chaplot, Casas, Bressand, Lengyel, Lample, Saulnier, Lavaud, Lachaux, Stock, Scao, Lavril, Wang, Lacroix, and Sayed]{jiang_mistral_2023}
Albert~Q. Jiang, Alexandre Sablayrolles, Arthur Mensch, Chris Bamford, Devendra~Singh Chaplot, Diego de~las Casas, Florian Bressand, Gianna Lengyel, Guillaume Lample, Lucile Saulnier, Lélio~Renard Lavaud, Marie-Anne Lachaux, Pierre Stock, Teven~Le Scao, Thibaut Lavril, Thomas Wang, Timothée Lacroix, and William~El Sayed.
\newblock Mistral {7B}, October 2023.
\newblock URL \url{http://arxiv.org/abs/2310.06825}.
\newblock arXiv:2310.06825 [cs].

\bibitem[Jiang et~al.(2025)Jiang, Lin, Cao, Tian, Kang, Wang, Sun, and Han]{jiang2025deepretrieval}
Pengcheng Jiang, Jiacheng Lin, Lang Cao, Runchu Tian, SeongKu Kang, Zifeng Wang, Jimeng Sun, and Jiawei Han.
\newblock Deepretrieval: Hacking real search engines and retrievers with large language models via reinforcement learning.
\newblock \emph{arXiv preprint arXiv:2503.00223}, 2025.

\bibitem[Jin et~al.(2025)Jin, Zeng, Yue, Yoon, Arik, Wang, Zamani, and Han]{jin_search-r1_2025}
Bowen Jin, Hansi Zeng, Zhenrui Yue, Jinsung Yoon, Sercan~O. Arik, Dong Wang, Hamed Zamani, and Jiawei Han.
\newblock Search-{R1}: {Training} {LLMs} to {Reason} and {Leverage} {Search} {Engines} with {Reinforcement} {Learning}.
\newblock August 2025.
\newblock URL \url{https://openreview.net/forum?id=Rwhi91ideu#discussion}.

\bibitem[Joachims et~al.(2017)Joachims, Granka, Pan, Hembrooke, and Gay]{joachims2017accurately}
Thorsten Joachims, Laura Granka, Bing Pan, Helene Hembrooke, and Geri Gay.
\newblock Accurately interpreting clickthrough data as implicit feedback.
\newblock In \emph{Acm Sigir Forum}, volume~51, pages 4--11. Acm New York, NY, USA, 2017.

\bibitem[Kurland and Tennenholtz(2022)]{compSearch}
Oren Kurland and Moshe Tennenholtz.
\newblock Competitive search.
\newblock In \emph{Proceedings of the 45th International ACM SIGIR Conference on Research and Development in Information Retrieval}, pages 2838--2849, 2022.

\bibitem[Lee et~al.(2024)Lee, Phatale, Mansoor, Mesnard, Ferret, Lu, Bishop, Hall, Carbune, Rastogi, and Prakash]{lee2024rlaifvsrlhfscaling}
Harrison Lee, Samrat Phatale, Hassan Mansoor, Thomas Mesnard, Johan Ferret, Kellie Lu, Colton Bishop, Ethan Hall, Victor Carbune, Abhinav Rastogi, and Sushant Prakash.
\newblock Rlaif vs. rlhf: Scaling reinforcement learning from human feedback with ai feedback, 2024.
\newblock URL \url{https://arxiv.org/abs/2309.00267}.

\bibitem[Li et~al.(2024)Li, Ding, Karten, and Jin]{li2024fightladder}
Wenzhe Li, Zihan Ding, Seth Karten, and Chi Jin.
\newblock Fightladder: A benchmark for competitive multi-agent reinforcement learning.
\newblock \emph{arXiv preprint arXiv:2406.02081}, 2024.

\bibitem[Madmon et~al.(2025{\natexlab{a}})Madmon, Pipano, Reinman, and Tennenholtz]{madmon2025nrd}
Omer Madmon, Idan Pipano, Itamar Reinman, and Moshe Tennenholtz.
\newblock On the convergence of no-regret dynamics in information retrieval games with proportional ranking functions.
\newblock In \emph{The Thirteenth International Conference on Learning Representations}, 2025{\natexlab{a}}.
\newblock URL \url{https://openreview.net/forum?id=jJXZvPe5z0}.

\bibitem[Madmon et~al.(2025{\natexlab{b}})Madmon, Pipano, Reinman, and Tennenholtz]{madmon2025search}
Omer Madmon, Idan Pipano, Itamar Reinman, and Moshe Tennenholtz.
\newblock The search for stability: Learning dynamics of strategic publishers with initial documents.
\newblock \emph{Journal of Artificial Intelligence Research}, 83, 2025{\natexlab{b}}.

\bibitem[Mao et~al.(2024)Mao, Cai, Xia, Wu, Wang, Wang, Ge, and Wei]{mao2024alympicsllmagentsmeet}
Shaoguang Mao, Yuzhe Cai, Yan Xia, Wenshan Wu, Xun Wang, Fengyi Wang, Tao Ge, and Furu Wei.
\newblock Alympics: Llm agents meet game theory -- exploring strategic decision-making with ai agents, 2024.
\newblock URL \url{https://arxiv.org/abs/2311.03220}.

\bibitem[Montazeralghaem et~al.(2020)Montazeralghaem, Zamani, and Allan]{RLRelevance2020}
Ali Montazeralghaem, Hamed Zamani, and James Allan.
\newblock A reinforcement learning framework for relevance feedback.
\newblock In \emph{Proceedings of the 43rd International ACM SIGIR Conference on Research and Development in Information Retrieval}, SIGIR '20, page 59–68, New York, NY, USA, 2020. Association for Computing Machinery.
\newblock ISBN 9781450380164.
\newblock \doi{10.1145/3397271.3401099}.
\newblock URL \url{https://doi.org/10.1145/3397271.3401099}.

\bibitem[Mordo et~al.(2025{\natexlab{a}})Mordo, Kordonsky, Nachimovsky, Tennenholtz, and Kurland]{mordo_lemss_2025}
Tommy Mordo, Tomer Kordonsky, Haya Nachimovsky, Moshe Tennenholtz, and Oren Kurland.
\newblock {LEMSS}: {LLM}-{Based} {Platform} for {Multi}-{Agent} {Competitive} {Search} {Simulation}.
\newblock In \emph{Proceedings of the 48th {International} {ACM} {SIGIR} {Conference} on {Research} and {Development} in {Information} {Retrieval}}, {SIGIR} '25, pages 3595--3605, New York, NY, USA, July 2025{\natexlab{a}}. Association for Computing Machinery.
\newblock ISBN 9798400715921.
\newblock \doi{10.1145/3726302.3730312}.
\newblock URL \url{https://dl.acm.org/doi/10.1145/3726302.3730312}.

\bibitem[Mordo et~al.(2025{\natexlab{b}})Mordo, Reinman, Tennenholtz, and Kurland]{mordo2025searchresultsdiversificationcompetitive}
Tommy Mordo, Itamar Reinman, Moshe Tennenholtz, and Oren Kurland.
\newblock Ameliorating the herding effect driven by search engines using diversity-based ranking.
\newblock In \emph{Proceedings of the 2025 International ACM SIGIR Conference on Innovative Concepts and Theories in Information Retrieval (ICTIR)}, ICTIR '25, page 1–11, New York, NY, USA, 2025{\natexlab{b}}. Association for Computing Machinery.
\newblock ISBN 9798400718618.
\newblock \doi{10.1145/3731120.3744600}.
\newblock URL \url{https://doi.org/10.1145/3731120.3744600}.

\bibitem[Nachimovsky and Tennenholtz(2025)]{nachimovsky2025power}
Haya Nachimovsky and Moshe Tennenholtz.
\newblock On the power of strategic corpus enrichment in content creation games.
\newblock In \emph{Proceedings of the AAAI Conference on Artificial Intelligence}, volume~39, pages 14019--14026, 2025.

\bibitem[Nachimovsky et~al.(2024)Nachimovsky, Tennenholtz, Raiber, and Kurland]{nachimovsky2024ranking}
Haya Nachimovsky, Moshe Tennenholtz, Fiana Raiber, and Oren Kurland.
\newblock Ranking-incentivized document manipulations for multiple queries.
\newblock In \emph{Proceedings of the 2024 ACM SIGIR International Conference on Theory of Information Retrieval}, pages 61--70, 2024.

\bibitem[Nachimovsky et~al.(2025)Nachimovsky, Tennenholtz, and Kurland]{nachimovsky2025multi}
Haya Nachimovsky, Moshe Tennenholtz, and Oren Kurland.
\newblock A multi-agent perspective on modern information retrieval.
\newblock \emph{arXiv preprint arXiv:2502.14796}, 2025.

\bibitem[Ouyang et~al.(2022)Ouyang, Wu, Jiang, Almeida, Wainwright, Mishkin, Zhang, Agarwal, Slama, Ray, et~al.]{ouyang2022training}
Long Ouyang, Jeffrey Wu, Xu~Jiang, Diogo Almeida, Carroll Wainwright, Pamela Mishkin, Chong Zhang, Sandhini Agarwal, Katarina Slama, Alex Ray, et~al.
\newblock Training language models to follow instructions with human feedback.
\newblock \emph{Advances in neural information processing systems}, 35:\penalty0 27730--27744, 2022.

\bibitem[Payal~Bajaj et~al.(2016)]{Bajaj2016Msmarco}
Nick~Craswell Payal~Bajaj, Daniel~Campos et~al.
\newblock Ms marco: A human generated machine reading comprehension dataset.
\newblock In \emph{InCoCo@NIPS}, 2016.

\bibitem[Peng et~al.(2023)Peng, Ding, Zhong, Shen, Liu, Zhang, Ouyang, and Tao]{peng2023towards_lang3}
Keqin Peng, Liang Ding, Qihuang Zhong, Li~Shen, Xuebo Liu, Min Zhang, Yuanxin Ouyang, and Dacheng Tao.
\newblock Towards making the most of {C}hat{GPT} for machine translation.
\newblock In Houda Bouamor, Juan Pino, and Kalika Bali, editors, \emph{Findings of the Association for Computational Linguistics: EMNLP 2023}, pages 5622--5633, Singapore, December 2023. Association for Computational Linguistics.
\newblock \doi{10.18653/v1/2023.findings-emnlp.373}.
\newblock URL \url{https://aclanthology.org/2023.findings-emnlp.373/}.

\bibitem[Qwen et~al.(2024)Qwen, Yang, Yang, Zhang, Hui, Zheng, Yu, Li, Liu, Huang, Wei, Lin, Yang, Tu, Zhang, Yang, Yang, Zhou, Lin, Dang, Lu, Bao, Yang, Yu, Li, Xue, Zhang, Zhu, Men, Lin, Li, Xia, Ren, Ren, Fan, Su, Zhang, Wan, Liu, Cui, Zhang, and Qiu]{qwen_qwen25_2024}
Qwen, An~Yang, Baosong Yang, Beichen Zhang, Binyuan Hui, Bo~Zheng, Bowen Yu, Chengyuan Li, Dayiheng Liu, Fei Huang, Haoran Wei, Huan Lin, Jian Yang, Jianhong Tu, Jianwei Zhang, Jianxin Yang, Jiaxi Yang, Jingren Zhou, Junyang Lin, Kai Dang, Keming Lu, Keqin Bao, Kexin Yang, Le~Yu, Mei Li, Mingfeng Xue, Pei Zhang, Qin Zhu, Rui Men, Runji Lin, Tianhao Li, Tingyu Xia, Xingzhang Ren, Xuancheng Ren, Yang Fan, Yang Su, Yichang Zhang, Yu~Wan, Yuqiong Liu, Zeyu Cui, Zhenru Zhang, and Zihan Qiu.
\newblock Qwen2.5 {Technical} {Report}, December 2024.
\newblock URL \url{http://arxiv.org/abs/2412.15115}.
\newblock arXiv:2412.15115 [cs].

\bibitem[Rafailov et~al.(2024)Rafailov, Sharma, Mitchell, Ermon, Manning, and Finn]{rafailov2024directpreferenceoptimizationlanguage}
Rafael Rafailov, Archit Sharma, Eric Mitchell, Stefano Ermon, Christopher~D. Manning, and Chelsea Finn.
\newblock Direct preference optimization: Your language model is secretly a reward model, 2024.
\newblock URL \url{https://arxiv.org/abs/2305.18290}.

\bibitem[Raifer et~al.(2017)Raifer, Raiber, Tennenholtz, and Kurland]{raifer2017information}
Nimrod Raifer, Fiana Raiber, Moshe Tennenholtz, and Oren Kurland.
\newblock Information retrieval meets game theory: The ranking competition between documents' authors.
\newblock In \emph{Proceedings of the 40th International ACM SIGIR Conference on Research and Development in Information Retrieval}, pages 465--474, 2017.

\bibitem[Raman et~al.(2024)Raman, Lundy, Amouyal, Levine, Leyton-Brown, and Tennenholtz]{raman2024steer}
Narun Raman, Taylor Lundy, Samuel Amouyal, Yoav Levine, Kevin Leyton-Brown, and Moshe Tennenholtz.
\newblock Steer: Assessing the economic rationality of large language models.
\newblock \emph{arXiv preprint arXiv:2402.09552}, 2024.

\bibitem[Raman et~al.(2025)Raman, Lundy, Amin, Perla, and Leyton-Brown]{raman2025steer}
Narun Raman, Taylor Lundy, Thiago Amin, Jesse Perla, and Kevin Leyton-Brown.
\newblock Steer-me: Assessing the microeconomic reasoning of large language models.
\newblock \emph{arXiv preprint arXiv:2502.13119}, 2025.

\bibitem[Rasley et~al.(2020)Rasley, Rajbhandari, Ruwase, and He]{10.1145/3394486.3406703}
Jeff Rasley, Samyam Rajbhandari, Olatunji Ruwase, and Yuxiong He.
\newblock Deepspeed: System optimizations enable training deep learning models with over 100 billion parameters.
\newblock In \emph{Proceedings of the 26th ACM SIGKDD International Conference on Knowledge Discovery \& Data Mining}, KDD '20, page 3505–3506, New York, NY, USA, 2020. Association for Computing Machinery.
\newblock ISBN 9781450379984.
\newblock \doi{10.1145/3394486.3406703}.
\newblock URL \url{https://doi.org/10.1145/3394486.3406703}.

\bibitem[Rathee et~al.(2025)Rathee, MacAvaney, and Anand]{rathee2025guiding}
Mandeep Rathee, Sean MacAvaney, and Avishek Anand.
\newblock Guiding retrieval using llm-based listwise rankers.
\newblock In \emph{European Conference on Information Retrieval}, pages 230--246. Springer, 2025.

\bibitem[Reimers and Gurevych(2019)]{reimers_sentence-bert_2019}
Nils Reimers and Iryna Gurevych.
\newblock Sentence-{BERT}: {Sentence} {Embeddings} using {Siamese} {BERT}-{Networks}, August 2019.
\newblock URL \url{http://arxiv.org/abs/1908.10084}.
\newblock arXiv:1908.10084 [cs].

\bibitem[Robertson et~al.(1993)Robertson, Walker, Jones, Hancock-Beaulieu, and Gatford]{robertson1993okapi}
S.~Robertson, S.~Walker, S.~Jones, M.~Hancock-Beaulieu, and M.~Gatford.
\newblock Okapi at trec-2.
\newblock In \emph{Proceedings of the TREC-2 Conference}, page 21–25, Gaithersburg, MD, 1993.

\bibitem[Schmied et~al.(2025)Schmied, Bornschein, Grau-Moya, Wulfmeier, and Pascanu]{unknown}
Thomas Schmied, Jörg Bornschein, Jordi Grau-Moya, Markus Wulfmeier, and Razvan Pascanu.
\newblock Llms are greedy agents: Effects of rl fine-tuning on decision-making abilities, 04 2025.

\bibitem[Schulman et~al.(2017)Schulman, Wolski, Dhariwal, Radford, and Klimov]{schulman2017proximalpolicyoptimizationalgorithms}
John Schulman, Filip Wolski, Prafulla Dhariwal, Alec Radford, and Oleg Klimov.
\newblock Proximal policy optimization algorithms, 2017.
\newblock URL \url{https://arxiv.org/abs/1707.06347}.

\bibitem[Shapira et~al.(2024{\natexlab{a}})Shapira, Madmon, Reichart, and Tennenholtz]{shapira2024can}
Eilam Shapira, Omer Madmon, Roi Reichart, and Moshe Tennenholtz.
\newblock Can llms replace economic choice prediction labs? the case of language-based persuasion games.
\newblock \emph{arXiv preprint arXiv:2401.17435}, 2024{\natexlab{a}}.

\bibitem[Shapira et~al.(2024{\natexlab{b}})Shapira, Madmon, Reinman, Amouyal, Reichart, and Tennenholtz]{shapira2024glee}
Eilam Shapira, Omer Madmon, Itamar Reinman, Samuel~Joseph Amouyal, Roi Reichart, and Moshe Tennenholtz.
\newblock Glee: A unified framework and benchmark for language-based economic environments.
\newblock \emph{arXiv preprint arXiv:2410.05254}, 2024{\natexlab{b}}.

\bibitem[Shapira et~al.(2025)Shapira, Madmon, Apel, Tennenholtz, and Reichart]{shapira2025human}
Eilam Shapira, Omer Madmon, Reut Apel, Moshe Tennenholtz, and Roi Reichart.
\newblock Human choice prediction in language-based persuasion games: Simulation-based off-policy evaluation.
\newblock 13:\penalty0 980--1006, 2025.
\newblock URL \url{https://doi.org/10.1162/TACL.a.16}.

\bibitem[Sharma et~al.(2022)Sharma, Patel, and Gupta]{sharma2022leveraging}
Amit Sharma, Neha Patel, and Rajesh Gupta.
\newblock Leveraging reinforcement learning and natural language processing for enhanced social media content optimization.
\newblock \emph{European Advanced AI Journal}, 11\penalty0 (8), 2022.

\bibitem[Shen et~al.(2023)Shen, Jin, Huang, Liu, Dong, Guo, Wu, Liu, and Xiong]{shen2023large}
Tianhao Shen, Renren Jin, Yufei Huang, Chuang Liu, Weilong Dong, Zishan Guo, Xinwei Wu, Yan Liu, and Deyi Xiong.
\newblock Large language model alignment: A survey.
\newblock \emph{arXiv preprint arXiv:2309.15025}, 2023.

\bibitem[Silver et~al.(2017)Silver, Schrittwieser, Simonyan, Antonoglou, Huang, Guez, Hubert, Baker, Lai, Bolton, Chen, Lillicrap, Hui, Sifre, van~den Driessche, Graepel, and Hassabis]{silver_mastering_2017}
David Silver, Julian Schrittwieser, Karen Simonyan, Ioannis Antonoglou, Aja Huang, Arthur Guez, Thomas Hubert, Lucas Baker, Matthew Lai, Adrian Bolton, Yutian Chen, Timothy Lillicrap, Fan Hui, Laurent Sifre, George van~den Driessche, Thore Graepel, and Demis Hassabis.
\newblock Mastering the game of {Go} without human knowledge.
\newblock \emph{Nature}, 550\penalty0 (7676):\penalty0 354--359, October 2017.
\newblock ISSN 1476-4687.
\newblock \doi{10.1038/nature24270}.
\newblock URL \url{https://www.nature.com/articles/nature24270}.
\newblock Publisher: Nature Publishing Group.

\bibitem[Silver et~al.(2018)Silver, Hubert, Schrittwieser, Antonoglou, Lai, Guez, Lanctot, Sifre, Kumaran, Graepel, Lillicrap, Simonyan, and Hassabis]{silver_general_2018}
David Silver, Thomas Hubert, Julian Schrittwieser, Ioannis Antonoglou, Matthew Lai, Arthur Guez, Marc Lanctot, Laurent Sifre, Dharshan Kumaran, Thore Graepel, Timothy Lillicrap, Karen Simonyan, and Demis Hassabis.
\newblock A general reinforcement learning algorithm that masters chess, shogi, and {Go} through self-play.
\newblock \emph{Science}, December 2018.
\newblock \doi{10.1126/science.aar6404}.
\newblock URL \url{https://www.science.org/doi/10.1126/science.aar6404}.
\newblock Publisher: American Association for the Advancement of Science.

\bibitem[Sun et~al.(2024)Sun, Liang, Yang, Xu, Yang, and Tong]{sun_rlrf4rec_2024}
Chao Sun, Yaobo Liang, Yaming Yang, Shilin Xu, Tianmeng Yang, and Yunhai Tong.
\newblock {RLRF4Rec}: {Reinforcement} {Learning} from {Recsys} {Feedback} for {Enhanced} {Recommendation} {Reranking}, October 2024.
\newblock URL \url{http://arxiv.org/abs/2410.05939}.

\bibitem[Susnjak(2024)]{susnjak2023applying_lang4}
Teo Susnjak.
\newblock Applying bert and chatgpt for sentiment analysis of lyme disease in scientific literature.
\newblock In \emph{Borrelia burgdorferi: Methods and Protocols}, pages 173--183. Springer, 2024.

\bibitem[Tennenholtz et~al.(2024)Tennenholtz, Chow, Hsu, Shani, Liang, and Boutilier]{tennenholtz_embedding-aligned_2024}
Guy Tennenholtz, Yinlam Chow, Chih-Wei Hsu, Lior Shani, Ethan Liang, and Craig Boutilier.
\newblock Embedding-{Aligned} {Language} {Models}, October 2024.
\newblock URL \url{http://arxiv.org/abs/2406.00024}.

\bibitem[Touvron et~al.(2023)Touvron, Lavril, Izacard, Martinet, Lachaux, Lacroix, Rozière, Goyal, Hambro, Azhar, Rodriguez, Joulin, Grave, and Lample]{touvron2023llamaopenefficientfoundation}
Hugo Touvron, Thibaut Lavril, Gautier Izacard, Xavier Martinet, Marie-Anne Lachaux, Timothée Lacroix, Baptiste Rozière, Naman Goyal, Eric Hambro, Faisal Azhar, Aurelien Rodriguez, Armand Joulin, Edouard Grave, and Guillaume Lample.
\newblock Llama: Open and efficient foundation language models, 2023.
\newblock URL \url{https://arxiv.org/abs/2302.13971}.

\bibitem[Vaswani et~al.(2017)Vaswani, Shazeer, Parmar, Uszkoreit, Jones, Gomez, Kaiser, and Polosukhin]{vaswani_attention_2017}
Ashish Vaswani, Noam Shazeer, Niki Parmar, Jakob Uszkoreit, Llion Jones, Aidan~N Gomez, Ł~ukasz Kaiser, and Illia Polosukhin.
\newblock Attention is {All} you {Need}.
\newblock In I.~Guyon, U.~Von Luxburg, S.~Bengio, H.~Wallach, R.~Fergus, S.~Vishwanathan, and R.~Garnett, editors, \emph{Advances in {Neural} {Information} {Processing} {Systems}}, volume~30. Curran Associates, Inc., 2017.
\newblock URL \url{https://proceedings.neurips.cc/paper_files/paper/2017/file/3f5ee243547dee91fbd053c1c4a845aa-Paper.pdf}.

\bibitem[Vinyals et~al.(2017)Vinyals, Ewalds, Bartunov, Georgiev, Vezhnevets, Yeo, Makhzani, K{\"u}ttler, Agapiou, Schrittwieser, et~al.]{vinyals2017starcraft}
Oriol Vinyals, Timo Ewalds, Sergey Bartunov, Petko Georgiev, Alexander~Sasha Vezhnevets, Michelle Yeo, Alireza Makhzani, Heinrich K{\"u}ttler, John Agapiou, Julian Schrittwieser, et~al.
\newblock Starcraft ii: A new challenge for reinforcement learning.
\newblock \emph{arXiv preprint arXiv:1708.04782}, 2017.

\bibitem[Vinyals et~al.(2019)Vinyals, Babuschkin, Czarnecki, Mathieu, Dudzik, Chung, Choi, Powell, Ewalds, Georgiev, et~al.]{vinyals2019grandmaster}
Oriol Vinyals, Igor Babuschkin, Wojciech~M Czarnecki, Micha{\"e}l Mathieu, Andrew Dudzik, Junyoung Chung, David~H Choi, Richard Powell, Timo Ewalds, Petko Georgiev, et~al.
\newblock Grandmaster level in starcraft ii using multi-agent reinforcement learning.
\newblock \emph{nature}, 575\penalty0 (7782):\penalty0 350--354, 2019.

\bibitem[von Werra et~al.(2020)von Werra, Belkada, Tunstall, Beeching, Thrush, Lambert, Huang, Rasul, and Gallouédec]{vonwerra2022trl}
Leandro von Werra, Younes Belkada, Lewis Tunstall, Edward Beeching, Tristan Thrush, Nathan Lambert, Shengyi Huang, Kashif Rasul, and Quentin Gallouédec.
\newblock Trl: Transformer reinforcement learning.
\newblock \url{https://github.com/huggingface/trl}, 2020.

\bibitem[Wang et~al.(2024{\natexlab{a}})Wang, Ma, Feng, Zhang, Yang, Zhang, Chen, Tang, Chen, Lin, Zhao, Wei, and Wen]{wang_survey_2024}
Lei Wang, Chen Ma, Xueyang Feng, Zeyu Zhang, Hao Yang, Jingsen Zhang, Zhiyuan Chen, Jiakai Tang, Xu~Chen, Yankai Lin, Wayne~Xin Zhao, Zhewei Wei, and Jirong Wen.
\newblock A survey on large language model based autonomous agents.
\newblock \emph{Frontiers of Computer Science}, 18\penalty0 (6):\penalty0 186345, March 2024{\natexlab{a}}.
\newblock ISSN 2095-2236.
\newblock \doi{10.1007/s11704-024-40231-1}.
\newblock URL \url{https://doi.org/10.1007/s11704-024-40231-1}.

\bibitem[Wang et~al.(2024{\natexlab{b}})Wang, Yang, Huang, Jiao, Yang, Jiang, Majumder, and Wei]{wang_text_2024}
Liang Wang, Nan Yang, Xiaolong Huang, Binxing Jiao, Linjun Yang, Daxin Jiang, Rangan Majumder, and Furu Wei.
\newblock Text {Embeddings} by {Weakly}-{Supervised} {Contrastive} {Pre}-training, February 2024{\natexlab{b}}.
\newblock URL \url{http://arxiv.org/abs/2212.03533}.
\newblock arXiv:2212.03533 [cs].

\bibitem[Wang et~al.(2024{\natexlab{c}})Wang, Li, Wang, Xing, Niu, Kong, Li, Long, Chang, and Zhang]{wang2024towards}
Qi~Wang, Jindong Li, Shiqi Wang, Qianli Xing, Runliang Niu, He~Kong, Rui Li, Guodong Long, Yi~Chang, and Chengqi Zhang.
\newblock Towards next-generation llm-based recommender systems: A survey and beyond.
\newblock \emph{arXiv preprint arXiv:2410.19744}, 2024{\natexlab{c}}.

\bibitem[Wang et~al.(2024{\natexlab{d}})Wang, Xie, Feng, Ding, Yang, and Xia]{wang2023chatgpt_lang5}
Zengzhi Wang, Qiming Xie, Yi~Feng, Zixiang Ding, Zinong Yang, and Rui Xia.
\newblock Is chat{GPT} a good sentiment analyzer?
\newblock In \emph{First Conference on Language Modeling}, 2024{\natexlab{d}}.
\newblock URL \url{https://openreview.net/forum?id=mUlLf50Y6H}.

\bibitem[Wu et~al.(2023)Wu, Zhu, Zhang, Wen, Ramchandran, and Jiao]{wu2023pairwiseproximalpolicyoptimization}
Tianhao Wu, Banghua Zhu, Ruoyu Zhang, Zhaojin Wen, Kannan Ramchandran, and Jiantao Jiao.
\newblock Pairwise proximal policy optimization: Harnessing relative feedback for llm alignment, 2023.
\newblock URL \url{https://arxiv.org/abs/2310.00212}.

\bibitem[Xenou et~al.(2018)Xenou, Chalkiadakis, and Afantenos]{xenou2018deep}
Konstantia Xenou, Georgios Chalkiadakis, and Stergos Afantenos.
\newblock Deep reinforcement learning in strategic board game environments.
\newblock In \emph{European Conference on Multi-Agent Systems}, pages 233--248. Springer, 2018.

\bibitem[Xi et~al.(2023)Xi, Chen, Guo, He, Ding, Hong, Zhang, Wang, Jin, Zhou, Zheng, Fan, Wang, Xiong, Zhou, Wang, Jiang, Zou, Liu, Yin, Dou, Weng, Cheng, Zhang, Qin, Zheng, Qiu, Huang, and Gui]{xi2023rise}
Zhiheng Xi, Wenxiang Chen, Xin Guo, Wei He, Yiwen Ding, Boyang Hong, Ming Zhang, Junzhe Wang, Senjie Jin, Enyu Zhou, Rui Zheng, Xiaoran Fan, Xiao Wang, Limao Xiong, Yuhao Zhou, Weiran Wang, Changhao Jiang, Yicheng Zou, Xiangyang Liu, Zhangyue Yin, Shihan Dou, Rongxiang Weng, Wensen Cheng, Qi~Zhang, Wenjuan Qin, Yongyan Zheng, Xipeng Qiu, Xuanjing Huang, and Tao Gui.
\newblock The rise and potential of large language model based agents: A survey, 2023.

\bibitem[Xie et~al.(2025)Xie, Rauch, and Zhang]{xie2025strategic}
Tian Xie, Pavan Rauch, and Xueru Zhang.
\newblock How strategic agents respond: Comparing analytical models with llm-generated responses in strategic classification.
\newblock \emph{arXiv preprint arXiv:2501.16355}, 2025.

\bibitem[Yang et~al.(2025)Yang, Penha, Palumbo, and Bouchard]{yang_aligned_2025}
Adam Yang, Gustavo Penha, Enrico Palumbo, and Hugues Bouchard.
\newblock Aligned {Query} {Expansion}: {Efficient} {Query} {Expansion} for {Information} {Retrieval} through {LLM} {Alignment}, July 2025.
\newblock URL \url{http://arxiv.org/abs/2507.11042}.
\newblock arXiv:2507.11042 [cs].

\bibitem[Yang et~al.(2016)Yang, Sloan, and Wang]{yang_dynamic_2016}
Grace~Hui Yang, Marc Sloan, and Jun Wang.
\newblock \emph{Dynamic information retrieval modeling}.
\newblock Number \#49 in Synthesis lectures on information concepts, retrieval, and services. Springer, Cham, Switzerland, 2016.
\newblock ISBN 978-3-031-01173-3 978-1-62705-526-0 978-3-031-02301-9.

\bibitem[Ye et~al.(2025)Ye, Xu, Sun, Xu, Wang, Dong, and Wen]{10.1145/3726302.3730026}
Xiaopeng Ye, Chen Xu, Zhongxiang Sun, Jun Xu, Gang Wang, Zhenhua Dong, and Ji-Rong Wen.
\newblock Llm-empowered creator simulation for long-term evaluation of recommender systems under information asymmetry.
\newblock In \emph{Proceedings of the 48th International ACM SIGIR Conference on Research and Development in Information Retrieval}, SIGIR '25, page 201–211, New York, NY, USA, 2025. Association for Computing Machinery.
\newblock ISBN 9798400715921.
\newblock \doi{10.1145/3726302.3730026}.
\newblock URL \url{https://doi.org/10.1145/3726302.3730026}.

\bibitem[Yuan et~al.(2023)Yuan, Yuan, Tan, Wang, Huang, and Huang]{yuan_rrhf_2023}
Zheng Yuan, Hongyi Yuan, Chuanqi Tan, Wei Wang, Songfang Huang, and Fei Huang.
\newblock {RRHF}: {Rank} {Responses} to {Align} {Language} {Models} with {Human} {Feedback} without tears, October 2023.
\newblock URL \url{http://arxiv.org/abs/2304.05302}.

\bibitem[Zhang et~al.(2023)Zhang, Liu, and Zhang]{zhang2023extractive_lang2}
Haopeng Zhang, Xiao Liu, and Jiawei Zhang.
\newblock Extractive summarization via {C}hat{GPT} for faithful summary generation.
\newblock In Houda Bouamor, Juan Pino, and Kalika Bali, editors, \emph{Findings of the Association for Computational Linguistics: EMNLP 2023}, pages 3270--3278, Singapore, December 2023. Association for Computational Linguistics.
\newblock \doi{10.18653/v1/2023.findings-emnlp.214}.
\newblock URL \url{https://aclanthology.org/2023.findings-emnlp.214/}.

\bibitem[Zhao et~al.(2024{\natexlab{a}})Zhao, Wang, Zhang, Jin, Zhu, Chen, and Xie]{zhao_competeai_2024}
Qinlin Zhao, Jindong Wang, Yixuan Zhang, Yiqiao Jin, Kaijie Zhu, Hao Chen, and Xing Xie.
\newblock {CompeteAI}: understanding the competition dynamics of large language model-based agents.
\newblock In \emph{Proceedings of the 41st {International} {Conference} on {Machine} {Learning}}, volume 235 of \emph{{ICML}'24}, pages 61092--61107, Vienna, Austria, July 2024{\natexlab{a}}. JMLR.org.

\bibitem[Zhao et~al.(2024{\natexlab{b}})Zhao, Liu, Ren, and Wen]{zhao2024dense}
Wayne~Xin Zhao, Jing Liu, Ruiyang Ren, and Ji-Rong Wen.
\newblock Dense text retrieval based on pretrained language models: A survey.
\newblock \emph{ACM Transactions on Information Systems}, 42\penalty0 (4):\penalty0 1--60, 2024{\natexlab{b}}.

\bibitem[Zhou et~al.(2024)Zhou, Agrawal, Zhang, Indurthi, Zhao, Song, Xu, and Zhu]{zhou_wpo_2024}
Wenxuan Zhou, Ravi Agrawal, Shujian Zhang, Sathish~Reddy Indurthi, Sanqiang Zhao, Kaiqiang Song, Silei Xu, and Chenguang Zhu.
\newblock {WPO}: {Enhancing} {RLHF} with {Weighted} {Preference} {Optimization}, October 2024.
\newblock URL \url{http://arxiv.org/abs/2406.11827}.

\bibitem[Zuze and Weideman(2013)]{zuze2013keyword}
Herbert Zuze and Melius Weideman.
\newblock Keyword stuffing and the big three search engines.
\newblock \emph{Online Information Review}, 37\penalty0 (2):\penalty0 268--286, 2013.

\end{thebibliography}

\appendix

\appendix

\section{Preliminaries}\label{app:pre}
\paragraph{Reinforcement Learning (RL)}
A Markov Decision Process (MDP) is defined as a tuple $(\mathcal{S}, \mathcal{A}, P, r, T, \gamma)$, where $\mathcal{S}$ is the state space, $\mathcal{A}$ is the action space, $P: \mathcal{S} \times \mathcal{A} \rightarrow \Delta(\mathcal{S})$ is the transition probability function, $r: \mathcal{S} \times \mathcal{A} \rightarrow \mathbb{R}$ is the reward function, $T$ is the episode horizon, and $\gamma \in [0, 1]$ is the discount factor. An agent interacts with the environment through a stationary stochastic policy $\pi : \mathcal{S} \rightarrow \Delta(\mathcal{A})$ that maps each state to a distribution over actions. The value of a policy $\pi$ at a state $s$ is the expected discounted return, defined as
\[
V^{\pi}(s) = \mathbb{E}_{P, \pi} \left[ \sum_{t=0}^{T-1} \gamma^t r(s_t, a_t) \,\middle|\, s_0 = s \right].
\]
The objective in reinforcement learning is to find an optimal policy $\pi^*$ that maximizes the expected value over an initial state distribution $\nu_0$, that is,
\[
\pi^* \in \arg\max_{\pi} \mathbb{E}_{s_0 \sim \nu_0} \left[ V^{\pi}(s_0) \right].
\]

\paragraph{Large Language Models (LLMs)}
A large language model (LLM) $\mathcal{L} : \mathcal{S} \mapsto \Delta_{\mathcal{S}}$ maps sequences of tokens to probability distributions over future sequences. These models are typically implemented using Transformer architectures~\citep{vaswani_attention_2017}, and are trained to predict the next token $x_t$ in a sequence, given the preceding tokens $(x_1, x_2, \ldots, x_{t-1})$, by minimizing the cross-entropy loss. Pre-trained LLMs vary significantly in size and capabilities, with larger models often exhibiting stronger reasoning, generalization, and generation performance. For example, the LLaMA 2 series~\citep{touvron2023llamaopenefficientfoundation} includes models with 7B, 13B, and 70B parameters.

\paragraph{Direct Preference Optimization (DPO)}
To align LLMs with human preferences, Direct Preference Optimization (DPO; ~\citealp{rafailov2024directpreferenceoptimizationlanguage}) provides a direct alternative to reinforcement learning methods such as PPO. DPO is usually trained on a dataset of human preferences in the form of tuples $(x, y_w, y_l)$, where $x$ is a prompt, $y_w$ is a preferred response, and $y_l$ is a less preferred one. Instead of using explicit reward modeling or rollout trajectories, DPO optimizes a contrastive loss that directly encourages the policy $\pi_\theta$ to assign higher likelihood to the preferred response relative to a reference policy $\pi_{\text{ref}}$. The DPO objective is defined as:

\[
\mathcal{L}_{\text{DPO}}(\pi_\theta; \pi_{\text{ref}}) = - \mathbb{E}_{(x, y_w, y_l) \sim \mathcal{D}} \left[
\log \sigma \left(
\beta \log \frac{\pi_\theta(y_w \mid x)}{\pi_{\text{ref}}(y_w \mid x)} -
\beta \log \frac{\pi_\theta(y_l \mid x)}{\pi_{\text{ref}}(y_l \mid x)}
\right)
\right],
\]

where $\sigma$ is the sigmoid function, $\beta > 0$ is a temperature parameter controlling the sharpness of the preference, $\pi_{\text{ref}}$ is typically set to the pre-trained base model and $D$ is a distribution over datapoints. This formulation introduces implicit regularization by comparing against the reference model and enables stable and efficient fine-tuning of LLMs using preference data, without requiring reinforcement learning rollouts or reward modeling.
\paragraph{A Schematic Figure of a Ranking Game}
\begin{figure}[H]
    \centering
    \includegraphics[width=0.40\linewidth]{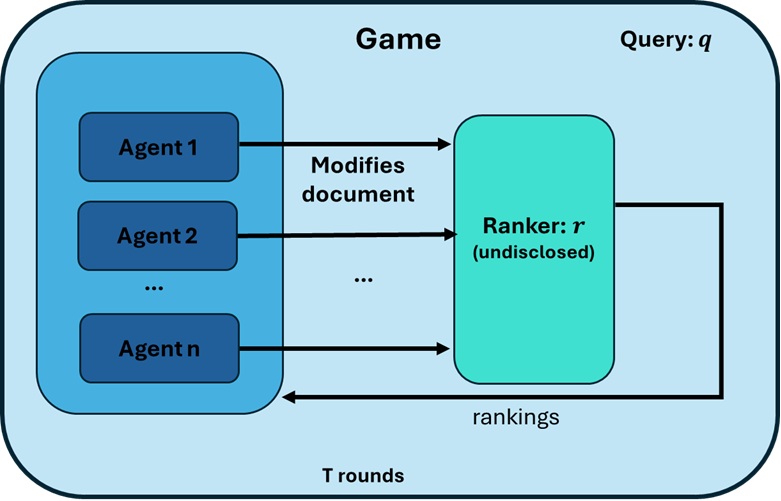}
        \caption{Illustration of a single game within a ranking competition. Each competition consists of multiple games. Each game is assigned with a query and composed of multiple rounds of agents' interaction. In each round, agents modify their documents and receive the rankings of each document.}
    \label{fig_scheme_game}
\end{figure}
\section{Data Generation}\label{app:data-gen}
\subsection{Static Generation (\staticDocsGeneration)}\label{app:data-gen-sg}
In this approach, we generate multiple relevant documents per query using an LLM prior to any optimization phase. We then collect the top-ranked and lowest-ranked documents (per query) to construct training triplets in the form (prompt, top-ranked document, lowest-ranked document).

We first prompt the LLM to generate a single relevant document for a given query, referred to as the \textit{pseudo-relevant document}. To ensure neutrality with respect to the ranking competition, we employ the instructional (system) prompt proposed by \cite{bardas_automatic_2025}, omitting any mention of competitive context. We then apply the Best-and-Worst-of-N (BWoN) sampling method, adapted from the Best-of-N strategy \citep{beirami2025theoreticalguaranteesbestofnalignment}. Given the pseudo-relevant document, we prompt the LLM $N$ times to generate $N$ \textit{modified documents} of the pseudo-relevant document. These documents are ranked using a ranking function. We collect the top-ranked and lowest-ranked documents from this set to construct training triplets in the form (prompt, top-ranked document, lowest-ranked document). This procedure is repeated for each query, yielding a dataset of preference pairs for downstream training. This method assumes access to a ranking function but no additional information about the competition dynamics or other participating agents. The prompts used to generate both the pseudo-relevant and the modified documents are presented in Figures \ref{no-feedback_prompt_example-init} and \ref{no-feedback_prompt_example}, respectively.

\begin{figure}[h!]
\centering
\input{prompts/no-feedback_prompt-init-document}
\caption{The prompt for generating the pseudo-relevant document.}
\label{no-feedback_prompt_example-init}
\end{figure}

\begin{figure}[h!]
\centering
\input{prompts/no-feedback_prompt}
\caption{The prompt for generating the modified documents with no past rankings feedback.}
\label{no-feedback_prompt_example}
\end{figure}

\subsection{Dynamic Generation (\dynamicDocsGeneration{})}\label{app:data-gen-dy}

While the static approach ignores the documents and rankings of other agents, the \dynamicDocsGeneration{} method explicitly models the dynamics of the competition. It does so by incorporating the documents and rankings of competing agents through simulations of repeated ranking games within the \testEnv{} environment \citep{mordo_lemss_2025}. We instantiate multiple copies of the LLM to simulate an $N$-player competition. The initial document in each training episode is generated by the LLM, following the same procedure as generating the pseudo-relevant document in \staticDocsGeneration{}. This choice, inspired by \citet{zhou_wpo_2024}, mitigates the off-policy distribution mismatch that can occur when the agent encounters states it has never seen during training; by ensuring the \rl{} learns from inputs representative of its training environment, we reduce instability and improve learning efficiency. In contrast, for evaluation we used initial documents drawn from a fixed dataset, ensuring that all agents received the same initial document for each query. This setup guarantees a common starting point and enables a fair comparison of strategies, following the standard approach adopted in prior work on competitive search \citep{raifer2017information, mordo2025searchresultsdiversificationcompetitive}. For each query and round selected from a set of rounds, we log the prompt presented to our agent, and extract the documents submitted by the highest- and lowest-ranked agents.

The resulting preference dataset consists of prompt-document triplets where the top and lowest ranked documents reflect actual competitive outcomes based on the simulated ranking environment. These data generation methods can be interpreted along the level of the agent's awareness of its downstream task and environment. As more contextual information becomes available, such as the identity or number of competing agents, the generated data increasingly approximates the true target distribution encountered during actual ranking competitions. Importantly, each method involves an inherent trade-off between exploration and sample efficiency: increasing the number of samples generated by the LLM can enhance exploratory coverage of the document space, thereby potentially improving the diversity of documents' scores of the resulting training data with respect to the ranking function. A systematic investigation of this trade-off is an important direction for future work.

\subsection{Parameters}
Document generation was performed using a temperature of $0.8$ to control sampling diversity \citep{yuan_rrhf_2023}. For both generation methods, we adopted the \nivList{} and \nivPair{} prompts \citep{bardas_automatic_2025}. In the \staticDocsGeneration{} method, we generated five modified documents per query and extracted training triplets consisting of the prompt, the top-ranked document, and the lowest-ranked document, based on a predefined ranking function. In the \dynamicDocsGeneration{} method, each simulated game involved five agents and lasted for $30$ rounds, following \cite{mordo_lemss_2025}. To match the dataset size of \staticDocsGeneration{}, we selected only one round per query: round $3$ for \nivList{} and round $4$ for \nivPair{}, as these are the first rounds with full ranking history required for the respective prompts.
\section{Hyper-parameters}
\label{appendix:hyperparams}

We report the hyper-parameters used in all generation and training phases.

\subsection{Generation Settings}\label{appendix:gen-hyper}
For all generative agents, we used the following decoding parameters during document generation:

\begin{itemize}
    \item \textbf{Temperature:} 0.8
    \item \textbf{Top-p (nucleus sampling):} 1.0 \hfill \textit{(as recommended in the TRL library)}
    \item \textbf{Top-k:} 0 \hfill \textit{(disables top-k filtering; used with top-p)}
\end{itemize}

\subsection{General Training Settings}\label{appendix:train}

We trained each LLM with two distinct datasets: \staticDocsGeneration{} and \dynamicDocsGeneration{}. Each LLM was fine-tuned on the 20 last transformer layers using the Transformer Reinforcement Learning (TRL) library \citep{vonwerra2022trl} and the DeepSpeed optimization framework \citep{10.1145/3394486.3406703}. Preliminary experiments indicated that fine-tuning fewer layers resulted in suboptimal performance, whereas deeper fine-tuning led to consistent improvements. We therefore selected 20 layers as a practical trade-off, given available resources. Due to computational constraints, some of training hyper-parameters were manually chosen with default values rather than tuned through extensive optimization. Our primary goal in this work is to establish and validate the alignment framework, rather than to exhaustively optimize agent performance. Nevertheless, as demonstrated in RQ1, even without hyper-parameter tuning, we successfully designed an \rl{} that outperforms the \prompt{}. We used the following optimization configuration:

\begin{itemize}
    \item \textbf{Batch size:} 2
    \item \textbf{Gradient accumulation steps:} 4
    \item \textbf{Number of epochs:} 4
    \item \textbf{Learning rate:} $1 \times 10^{-6}$
    \item \textbf{Number of trainable transformer layers:} 20
    \item \textbf{Loss:} WPO (weighted DPO variant; \citealp{zhou_wpo_2024})
    \item \textbf{DPO/WPO beta:} 0.1
\end{itemize}
We used the Adam optimizer with the following configuration:
\begin{itemize}
    \item \textbf{Beta 1:} 0.9
    \item \textbf{Beta 2:} 0.99
    \item \textbf{Weight decay:} 0.01
\end{itemize}

\section{Evaluation Measures}\label{appendix:eval}

\paragraph{Scaled Promotion}  
To quantify how effectively a document modification improves ranking within a single round, we use the \textit{Scaled Promotion} metric. It measures the normalized improvement (or demotion) in rank between consecutive rounds:

\begin{equation}
\text{Scaled Promotion$_t$}(d) = \frac{\text{Rank}_{t}(d) - \text{Rank}_{t+1}(d)}{\max\big(\text{Rank}_{t}(d)-1, \; N - \text{Rank}_{t}(d)\big)}
\end{equation}

where $\text{Rank}_{t}(d)$ is the rank of document $d$ in round $t$, $\text{Rank}_{t+1}(d)$ is its rank in the following round, and $N$ is the number of competing documents. The denominator represents the maximum achievable promotion (if the document is not ranked first) or demotion (if it is not ranked last). A higher score indicates a stronger relative promotion, normalized by what is theoretically possible.

\paragraph{OrigFaith (faithfulness to the original document)}  
Given an original (initial) document $d_{\text{orig}}$ and a modified document 
$d_{\text{mod}} = \{s_1, \dots, s_m\}$ with $m$ sentences, we first compute the
\textit{Raw Faithfulness} score using an NLI-based model (TrueTeacher, TT) \citep{gekhman2023trueteacherlearningfactualconsistency}:

\begin{equation}
\text{RawFaith}(d_{\text{mod}}, d_{\text{orig}})
= \frac{1}{m} \sum_{i=1}^m 
\mathbbm{1}\bigl\{ \text{TT}(s_i, d_{\text{orig}}) \geq 0.5 \bigr\}.
\end{equation}

where $\text{TT}(s_i, d_{\text{orig}}) \in [0,1]$ is the entailment probability
between the modified sentence $s_i$ and the original document $d_{\text{orig}}$,
and $0.5$ is a predefined entailment threshold chosen according to \citet{gekhman2023trueteacherlearningfactualconsistency}.  

To account for varying document lengths and ensure comparability across
instances, we normalize the RawFaith score:

\begin{equation}
\text{OrigFaith}(d_{\text{mod}}, d_{\text{orig}}) \;=\; 
\frac{\text{RawFaith}(d_{\text{mod}}, d_{\text{orig}})}{\text{RawFaith}(d_{\text{orig}}, d_{\text{orig}})},
\end{equation}

This yields a normalized faithfulness score in $[0,1]$ that reflects how well the modified
document preserves the faithfulness to the original document.

\paragraph{Win-rate}  
This metric measures how frequently an agent achieves the top rank across rounds, averaged over all queries (games). It is defined as:

\begin{equation}
\text{Win Rate} = \frac{1}{|Q|} \sum_{q=1}^{|Q|} \frac{W_q}{R_q}
\end{equation}

where $|Q|$ is the number of queries in the evaluation set, $W_q$ is the number of rounds in which the agent ranked first for query $q$, and $R_q$ is the total number of rounds played for query $q$. This metric captures the agent’s ability to consistently produce top-ranked outputs relative to its competitors. We report the win-rate of the \rl{} and compare it against two baselines: (i) a random baseline, equal to $\frac{1}{\text{\#players}}$, and (ii) the \prompt{} with the best performance with respect to the win-rate.

\section{Robustness of the \rl{} Performance}
\label{ape:robust}

We evaluate the robustness of the \rl{} with respect to two key competition parameters: (1) the number of competing agents, and (2) the sampling temperature of the agent’s LLM at evaluation. Studying these aspects is crucial for understanding whether a trained agent remains effective when deployed under varying and potentially unpredictable conditions. For example, in practical environments, the number of competitors and the behavior of LLM-based agents (e.g., due to randomness introduced by sampling) may fluctuate significantly. Thus, an agent’s resilience to such changes is an important factor in its practical utility.
We focus on the best-performing agent from RQ1 (See Section \ref{sec:rq1}.): a Mistral-based model trained using the \dynamicDocsGeneration{} procedure with \nivList{} prompting. For both training and evaluation, we use the E5-unsupervised ranker \citep{wang_text_2024}, which demonstrated superior performance in past work over other ranking functions, and has also been adopted in prior work on competitive search \citep{mordo2025searchresultsdiversificationcompetitive,bardas_automatic_2025}.

Table \ref{tab:rq4} reports the win rates of our agent in competitions with $1$, $4$, and $7$ competitors. The results are presented for the \homoEval{} setting, in which each competitor is a duplication instance of the same \prompt{}. We did not consider the \hetroEval{} setting in order to isolate the effect of the number of agents from potential confounding factors related to the choice of language model.
We evaluate the agent at two sampling temperatures: $0.5$ and $1.0$. Temperature $0.0$, used in previous RQs, is omitted here as it prevents exploration of stochastic behavior in competitive settings. Across all configurations, the \rl{} consistently outperforms the best \prompt{}. As expected, the win-rate decreases with the number of competitors due to increased competition, but remains significantly above the random baseline.

Table \ref{tab:rq42} presents the results of a broader temperature sweep, evaluating the agent at temperatures $0.0$, $0.5$, $0.8$ (matching the temperature used during data generation), $1.0$, $1.5$, and $2.0$. We fixed the number of competitors as five, under the \hetroEval{} setting. In all tested temperatures, our agent maintains a win-rate in the range $[0.58, 0.62]$, significantly outperforming all competitors across the board. These findings demonstrate that the \rl{} is robust to variation in both the number of competitors and the temperature at the evaluation phase.


\begin{table}[t]
\centering
\small
\caption{Comparison of performances in \homoEval{} competitions with Mistral 8B agents trained with \dynamicDocsGeneration{} and prompted with \nivList{} under different number of \prompt{}s compete the \rl{}. We report the win-rate of the \rl{} and the best \prompt{}. '*' marks a statistically significant difference with the win-rate of the best \prompt{}. The best performance in each configuration is boldfaced.}

\begin{tabular}{llcc|cc}
\toprule
\textbf{LLM} & \textbf{Train Setting} & \textbf{Temp.} & \textbf{\# \prompt{}s} 
& \makecell[c]{\textbf{\rl{}}\\\textbf{WR}} & \makecell[c]{\textbf{\prompt{}}\\\textbf{WR}} \\
\midrule
\multirow{6}{*}{Mistral} & 
\multirow{6}{*}{\dynamicDocsGeneration{} (\nivList{})} & 
\multirow{3}{*}{$0.5$} & $1$ & $\mathbf{0.72^{*}}$ & $0.28$ \\
&&& $4$ & $\mathbf{0.65^{*}}$ & $0.11$ \\
&&& $7$ & $\mathbf{0.65^{*}}$ & $0.06$ \\
\cmidrule(l){3-6}
&& \multirow{3}{*}{$1$} & $1$ & $\mathbf{0.74^{*}}$ & $0.26$ \\
&&& $4$ & $\mathbf{0.60^{*}}$ & $0.11$ \\
&&& $7$ & $\mathbf{0.58^{*}}$ & $0.07$ \\
\bottomrule
\end{tabular}

\label{tab:rq4}
\end{table}

\begin{table}[t]
\centering
\small
\caption{Comparison of performances in \hetroEval{} competitions with Mistral 8B agents trained with \dynamicDocsGeneration{} and prompted with \nivList{} under temperatures of the LLM at evaluation time. We report the win-rate of the \rl{} and the best \prompt{}. '*' marks a statistically significant difference with the win-rate of the best \prompt{}. The best performance in each configuration is boldfaced.}

\begin{tabular}{llc|cc}
\toprule
\textbf{LLM} & \textbf{Train Setting} & \textbf{Temp.} 
& \makecell[c]{\textbf{\rl{}}\\\textbf{WR}}  & \makecell[c]{\textbf{Best \prompt{}}\\\textbf{WR}} \\
\midrule
\multirow{5}{*}{Mistral} & 
\multirow{5}{*}{\dynamicDocsGeneration{} (\nivList{})} & $0.5$  & $\mathbf{0.60^{*}}$ & $0.11$ \\
&& $0.8$  & $\mathbf{0.58^{*}}$ & $0.11$ \\
&& $1$    & $\mathbf{0.62^{*}}$ & $0.11$ \\
&& $1.5$  & $\mathbf{0.62^{*}}$ & $0.15$ \\
&& $2$    & $\mathbf{0.58^{*}}$ & $0.11$ \\
\bottomrule
\end{tabular}

\label{tab:rq42}
\end{table}

\section{Effectiveness of the \rl{} in single-round offline evaluation}\label{app:rq3new}


We evaluate the \rl{} and compare its performance to that of \prompt{}s in the single-round setting introduced by \citet{bardas_automatic_2025}. In this setting, each agent modifies documents in the context of an existing competition previously conducted between students \citep{mordo2025searchresultsdiversificationcompetitive}; the students were rewarded to improve their rankings. We consider an \rl{} with Mistral 8B language model with the \nivList{} prompt, trained using the \dynamicDocsGeneration{} procedure with the E5-unsupervised ranking function; the same ranker was used in ranking competitions with human participants \citep{mordo2025searchresultsdiversificationcompetitive}. The \prompt{} used also the \nivList{} prompt.

We report two evaluation measures: scaled promotion and faithfulness. The scaled promotion metric is used to quantify ranking properties, computed per player and her document for a query. Specifically, it measures the change in a document’s rank between consecutive rounds, defined as the number of positions by which the document is promoted (or demoted), normalized by the maximum potential promotion (or demotion) given the document’s position. The values for the students are averaged over them and the queries, while the values for an agent is averaged over queries. The faithfulness\footnote{See Section \ref{sec:eval}.} captures whether the modifications preserve the factual consistency of the original document; it is measured using the NLI-based approach proposed by \citet{gekhman2023trueteacherlearningfactualconsistency}. Formal definitions are presented in Appendix \ref{appendix:eval}.

Table \ref{tab:rq1} shows that among the agents with Mistral 8B, the \rl{} achieves scaled promotion that is higher than that of the student\footnote{Note that student scores vary across rows as they depend on the agent under evaluation.} participants and the \prompt{}. Additionally, the \rl{} also demonstrates higher faithfulness to the original document than the \prompt{}.
However, both the RL and \prompt{}s exhibit faithfulness scores lower than those of human participants. This indicates that while LLM-based agents are effective at strategic promotion, they may struggle to preserve content faithfulness relative to human baselines.

Additionally, the scaled promotion and faithfulness scores of the RL and \prompt{}s are lower than those reported by \citet{bardas_automatic_2025} for \gpt{}-based agents. This performance gap is attributed to the use of language models with 8B parameters in our experiments, a significantly smaller model compared to \gpt{}. To support this claim, Table \ref{tab:rq12} includes results for \prompt{}s with larger language models: Llama 70B, Qwen2.5 32B and Gemma2 27B which indeed outperform both the Mistral 8B variants in both scaled promotion and faithfulness. In future research we intend to explore training methods to optimize not only the ranking promotion but also the faithfulness to the initial document.


\begin{table}[t]
\caption{Performance comparison of Mistral 8B \rl{} and \prompt{} using the \nivList{} prompt. The \rl{} was trained with the \dynamicDocsGeneration{} procedure. The Table presents scaled promotion and faithfulness scores from a single-round offline evaluation conducted on an existing ranking competition following the setup of \citet{bardas_automatic_2025}.}
\centering
\small

\begin{tabular}{l|cc|cc|cc|cc|cc}
\toprule
\textbf{LLM} & 
\multicolumn{2}{c|}{\textbf{Scaled Promotion}} &
\multicolumn{2}{c|}{\textbf{Faithfulness}} \\
\cmidrule(lr){2-3} \cmidrule(lr){4-5}
& \textbf{Students} & \textbf{The Agent} 
& \textbf{Students} & \textbf{The Agent} \\
\midrule
Mistral 8B + RL
& $0.089$ & $0.266$ 
& $0.788$ & $0.408$ \\

Mistral 8B
& $0.328$ & $-0.363$ 
& $0.788$ & $0.350$ \\
\bottomrule
\end{tabular}

\label{tab:rq1}
\end{table}

\begin{table}[t]
\caption{Performance comparison of \prompt{}s with larger (than 8B) language models. The table presents scaled promotion and faithfulness scores from a single-round offline evaluation conducted on an existing ranking competition following the setup of \citet{bardas_automatic_2025}.}
\centering
\small
\begin{tabular}{l|cc|cc|cc|cc|cc}
\toprule
\textbf{LLM} & 
\multicolumn{2}{c|}{\textbf{Scaled Promotion}} &
\multicolumn{2}{c|}{\textbf{Faithfulness}} \\
\cmidrule(lr){2-3} \cmidrule(lr){4-5}
& \textbf{Students} & \textbf{The Agent} 
& \textbf{Students} & \textbf{The Agent} \\
\midrule
Llama 70B 
& $0.100$ & $0.270$ 
& $0.788$ & $0.785$ \\

Qwen2.5 32B 
& $0.226$ & $-0.119$ 
& $0.788$ & $0.666$ \\

Gemma2 27B 
& $0.086$ & $0.296$ 
& $0.788$ & $0.936$ \\
\bottomrule
\end{tabular}
\label{tab:rq12}
\end{table}

\section{Analysis of Strategies}\label{app:strategies}
To complement the win-rate results (See Section \ref{sec:rq1}), we analyze the underlying strategies that the \rl{} and \prompt{} employ when modifying their documents over time. Our focus is on the settings with Mistral from RQ1: (i) competitions with the \nivList{} prompt under \dynamicDocsGeneration{}, evaluated in both the \homoEval{} and \hetroEval{} settings, and (ii) the \staticDocsGeneration{} generation method under the \hetroEval{} setting.

We adopt several measures introduced by \citet{mordo_lemss_2025}, chosen to capture both player-level and ranked-list-level dynamics. First, we measured the \textit{diversity} of documents by computing the minimum inter-document similarity within a ranked list across rounds. This measure reflects how varied the documents remain throughout the competition. Second, we evaluated the \textit{convergence} of a competition at the player-level. We measured the similarity between documents produced by the same agent between consecutive rounds. This indicates the extent to which agents continue modifying their documents as the competition progresses, and whether their strategies stabilize over time. Third, we track the \textit{scores} assigned by the ranking function to the documents of both the \rl{} and the \prompt{}s. For the \prompt{}s, we arbitrarily selected one representative per competition.

\begin{figure}[ht!]
    \centering
    \includegraphics[scale=0.4]{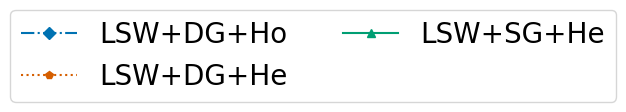}
    \begin{minipage}[t]{0.8\textwidth}
        \centering
        \includegraphics[width=\linewidth]{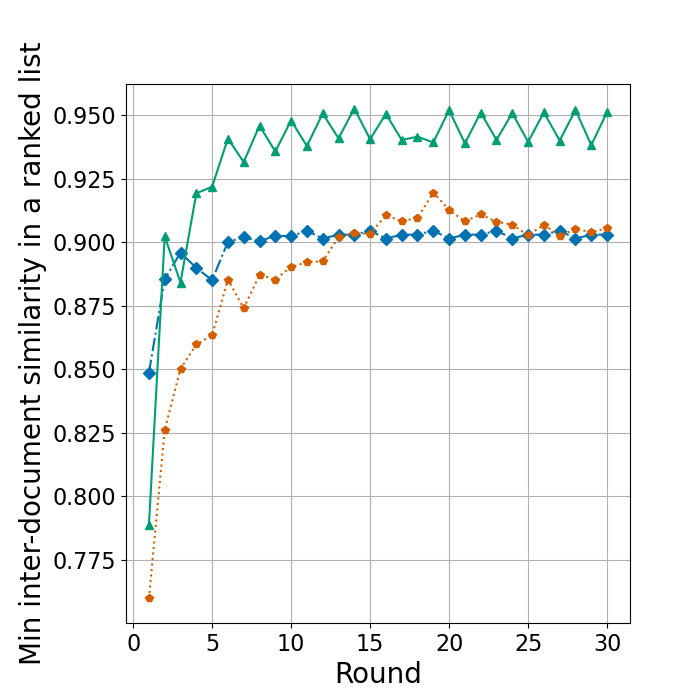}
    \end{minipage}
    \caption{Comparison of the average minimum inter-document similarity in a ranked list across rounds, for the settings in RQ1: (i) \dynamicDocsGeneration{} with the \nivList{} prompt under the \homoEval{} setting, (ii) \dynamicDocsGeneration{} with the \nivList{} prompt under the \hetroEval{} setting, and (iii) \staticDocsGeneration{} under the \hetroEval{} setting.}
    \label{fig_anal_1}
\end{figure}

\begin{figure}[ht!]
\includegraphics[scale=0.4]{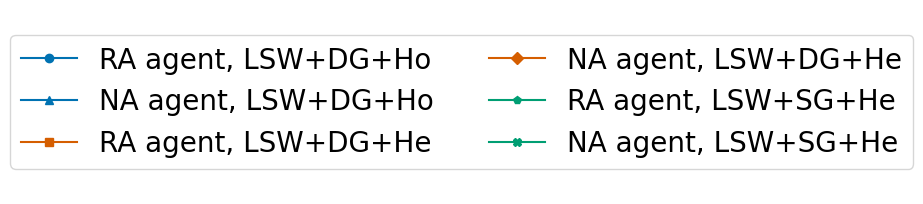}
    \centering
        \begin{minipage}[t]{0.48\textwidth}
        \centering
        \includegraphics[width=\linewidth]{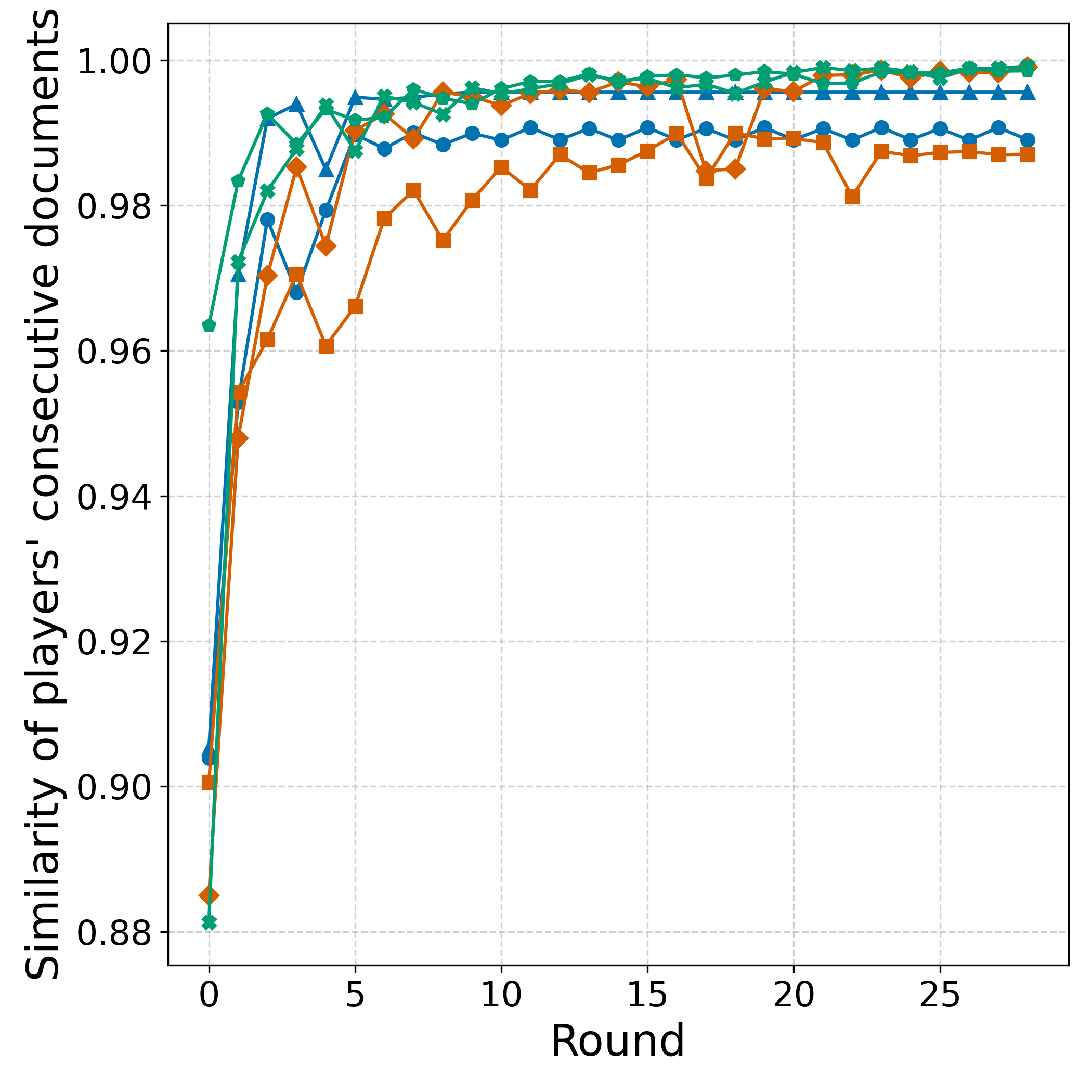}
        \subcaption[]{}{}
        \label{fig_anal_3}
    \end{minipage}
    \hfill
        \begin{minipage}[t]{0.48\textwidth}
        \centering
        \includegraphics[width=\linewidth]{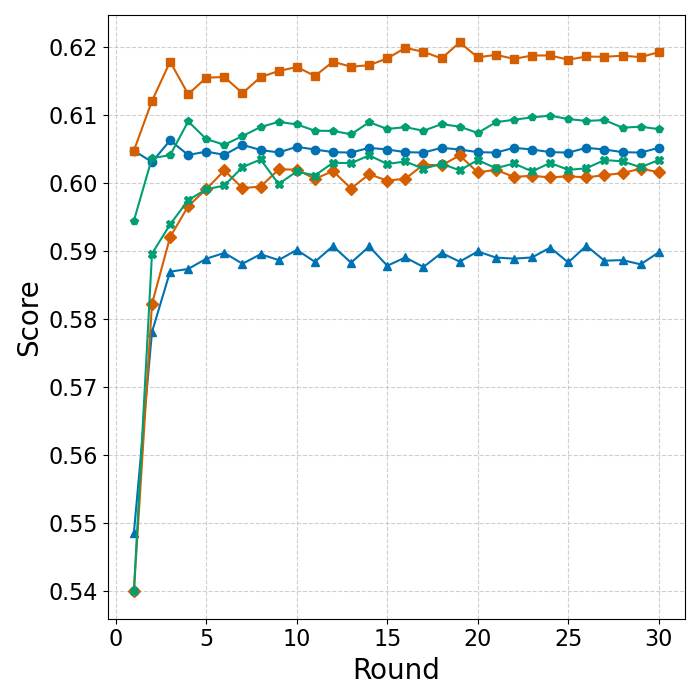}
        \subcaption[]{}{}
        \label{fig_anal_4}
    \end{minipage}
    \caption{Comparison of the RL-aligned agent (\textbf{RA agent}) and non-aligned agents (\prompt{}s) under the RQ1 settings: (i) \dynamicDocsGeneration{} with the \nivList{} prompt under the \homoEval{} setting, (ii) \dynamicDocsGeneration{} with the \nivList{} prompt under the \hetroEval{} setting, and (iii) \staticDocsGeneration{} under the \hetroEval{} setting. We evaluate the following measures: (a) the average similarity (over queries) between consecutive rounds ($i, i+1$), and (b) the average ranking score over rounds (averaged over queries).}
    \label{fig_anal_2}
\end{figure}

To compute similarity measures, we use S-BERT \citep{reimers_sentence-bert_2019} as the encoder for document representations and apply cosine similarity to their embeddings. The results as a function of the round are presented in Figures \ref{fig_anal_1} and \ref{fig_anal_2}.

Figure \ref{fig_anal_1} shows that the minimum inter-document similarity is consistently higher under the \staticDocsGeneration{} setup than under \dynamicDocsGeneration{}. This is expected, since in the static case the agent was trained on a self-generated dataset independent of competitive dynamics, which tends to reduce variation and increase homogeneity across the ranked list. In contrast, the dynamic setup relies on preference data derived from competitions between Mistral clones. This training process exposes the agent to a broader range of document modifications, ultimately fostering greater diversity in the ranked lists.

Figure \ref{fig_anal_3} examines the similarity of each agent’s consecutive documents. The \rl{} trained with \dynamicDocsGeneration{} (in both \homoEval{} and \hetroEval{} settings) display the lowest similarity between rounds, indicating that they adapt their documents more substantially across iterations. \prompt{}s, by contrast, exhibit more conservative and homogeneous modifications. In all settings, the \rl{} and \prompt{}s demonstrate a tendency to converge toward stable strategies. This convergence is consistent with the herding effect observed in ranking competitions between LLMs \citep{mordo_lemss_2025}, where agents gradually reduce exploration and adopt increasingly similar behaviors.

Finally, Figure \ref{fig_anal_4} plots the ranking function’s score assigned to documents over rounds. Across all settings, the \rl{} achieves higher scores than the \prompt{}, reflecting the alignment induced by RLRF training. Notably, \prompt{}s start with relatively low scores but quickly improve during the first few rounds before stabilizing at a plateau. The \rl{}, however, is already aligned to the ranking function at the outset, and thus shows smaller relative gains during the competition.

We now turn to analyze the win-rates with respect to the first and last rounds. The first round reflects the initial alignment of the \rl{}, while the last round (round 30) captures the dynamics that unfold during the competition. Table \ref{tab:wr-rounds} reports the win rates of the \rl{} and the \prompt{} in the first and last rounds. In round 1, the \rl{} consistently and significantly outperformed the \prompt{} across all three settings, demonstrating the effectiveness of its alignment procedure. By round 30, the \prompt{} had improved its win rate in two of the three settings, yet the \rl{} still maintained a clear advantage. This improvement of the \prompt{} is consistent with the herding effect, whereby agents converge toward similar strategies over repeated rounds. Overall, the results show that the alignment process benefits the \rl{} in two ways: it enhances its alignment with the ranking function and strengthens its ability to compete against opponents during the ranking competition.

\begin{table}[t]
\centering
\caption{Win-rate (WR) of the \rl{} and the \prompt{} at the first and last rounds ($1$,$30$) across the configurations and agents. $r$ denotes statistical significance difference between rounds ($01$ vs. $30$) for the same player and setting. 
$p$ denotes statistical significance difference between the NA and \rl{}s at the same round and setting.}
\begin{tabular}{l|cc|cc}
\toprule
\multirow{2}{*}{Configuration} & \multicolumn{2}{c|}{\prompt{}} & \multicolumn{2}{c|}{\rl{}} \\
\cmidrule(lr){2-3} \cmidrule(lr){4-5}
 & Round $01$ & Round $30$ & Round $01$ & Round $30$ \\
\midrule
\nivList{}+\staticDocsGeneration{}+\hetroEval{} & $0.12$ & $0.20$ & $0.50^{p,r}$ & $0.22$ \\
\nivList{}+\dynamicDocsGeneration{}+\homoEval{}   & $0.02$ & $0.14$ & $0.88^{p}$ & $0.72^{p}$ \\
\nivList{}+\dynamicDocsGeneration{}+\hetroEval{} & $0.06$ & $0.06$ & $0.68^{p}$ & $0.66^{p}$ \\
\bottomrule
\end{tabular}

\label{tab:wr-rounds}
\end{table}

\section{Multiple \rl{}s}\label{app:multi-rlrf}
We study the effect of the participation of multiple \rl{}s in ranking competitions. We focus on \rl{}s based on Mistral language model, trained using \dynamicDocsGeneration{} and prompted with \nivList{}. All competitions involve five agents with Mistral language models (\rl{} and \prompt{}s). In contrast to RQ1, where we used the same LLM hyper-parameters as \citet{bardas_automatic_2025}, here we set the temperature to $1$ (instead of $0$) to increase the dynamics of document generation. In each setting, we increment the number of \rl{}s by one while decreasing the number of \prompt{}s accordingly. We report the same measures as in Appendix \ref{app:strategies}.

Figure \ref{fig_multirl_5} shows the minimum inter-document similarity in ranked lists across rounds. In all settings, the similarity in a ranked list increases over rounds, consistent with prior work \citep{mordo_lemss_2025}. In addition, increasing the number of \rl{}s generally leads to higher minimum inter-document similarity, contrasting with the single \rl{} scenario (Figure \ref{fig_anal_3}), where the \rl{} modifies its documents more extensively than \prompt{}s. This suggests that alternative modification strategies emerge when multiple \rl{}s compete.

\begin{figure}[ht!]
    \centering
    
    \begin{minipage}{0.4\textwidth}
        \centering
        \includegraphics[width=\linewidth]{figs/MultiRLRF/graphs_comparison.png}
    \end{minipage}
    
    
    \begin{minipage}[t]{0.45\textwidth}
        \centering
        \includegraphics[width=\linewidth]{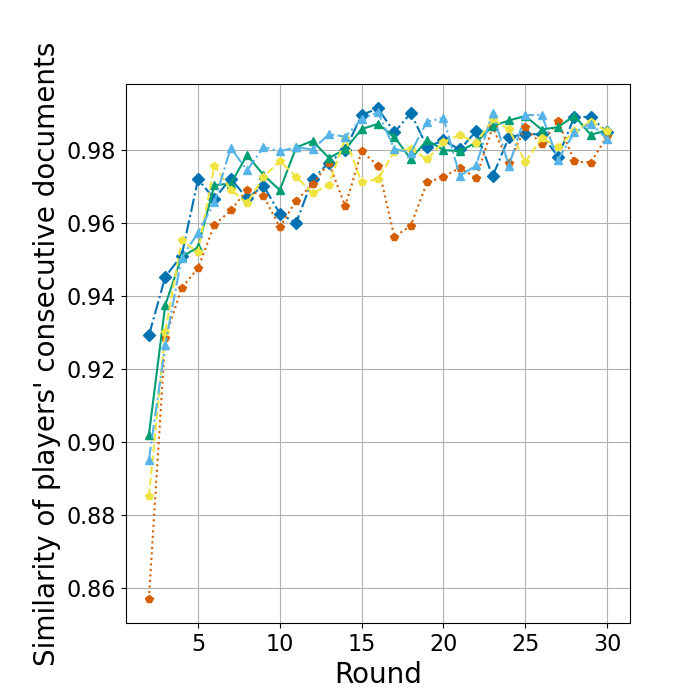}
        \subcaption[]{}{}
        \label{fig_multirl_6}
    \end{minipage}
    \hfill
    \begin{minipage}[t]{0.45\textwidth}
        \centering
        \includegraphics[width=\linewidth]{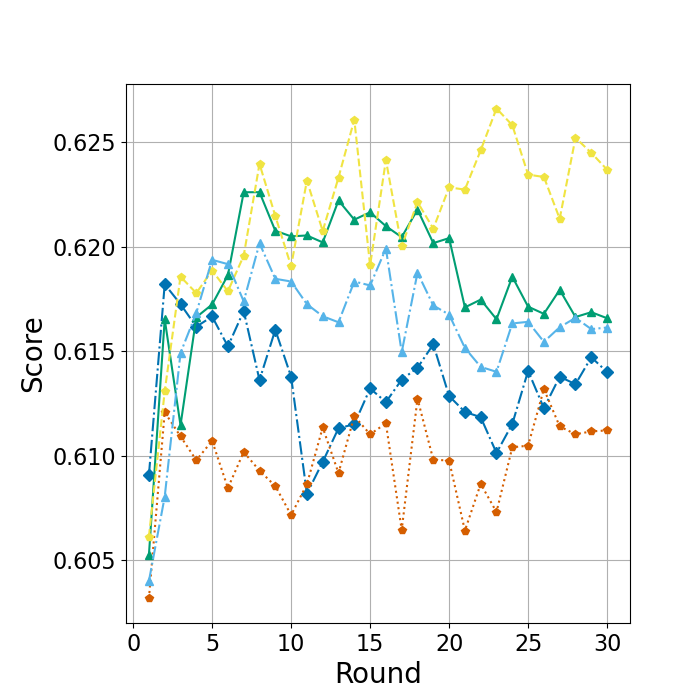}
        \subcaption[]{}{}
        \label{fig_multirl_7}
    \end{minipage}
    \caption{Comparison of the \rl{} and \prompt{}s across settings with varying numbers of \rl{}s. In each setting, five agents compete: the number of \rl{}s ranges from one to five, and the remaining agents are \prompt{}s. Each setting with $j$ \rl{}s is abbreviated as $j$ \textbf{RA agent}. We evaluate the following measures: (a) the average similarity (over queries) between consecutive rounds ($i, i+1$), and (b) the average ranking score over rounds (averaged over queries).}
    \label{fig_anal_8}
\end{figure}

Figure \ref{fig_multirl_6} shows the similarity between consecutive documents of the \rl{}s across settings. In all settings, similarity increases over rounds, consistent with prior work on ranking competitions between LLMs \citep{mordo_lemss_2025}. No clear differences are observed between settings. A possible explanation is that the presence of multiple \rl{}s stabilizes document modification strategies across settings. Figure \ref{fig_multirl_7} presents the ranking scores across rounds, which exhibit a slight upward trend without statistically significant differences between settings.

Overall, these results indicate that introducing multiple \rl{}s influences document modification dynamics, increasing similarity between ranked documents while maintaining consistent ranking performance across rounds.
\begin{figure}[ht!]
    \centering
    \includegraphics[scale=0.4]{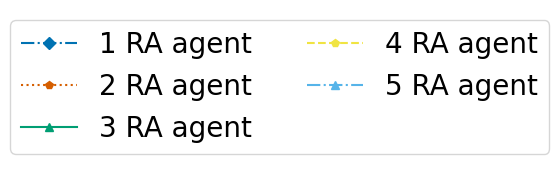}
    \begin{minipage}[t]{0.8\textwidth}
        \centering
        \includegraphics[width=\linewidth]{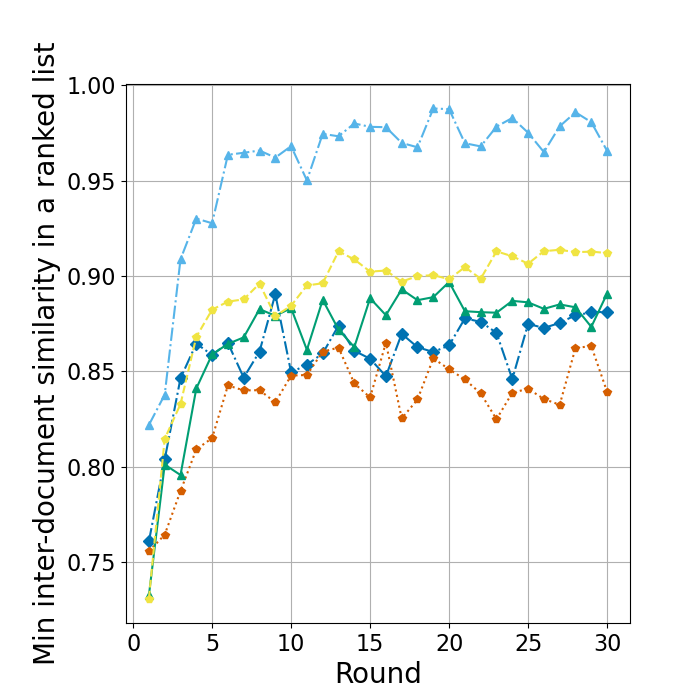}

    \end{minipage}
    \caption{Comparison of the average minimum inter-document similarity in a ranked list (averaged across rounds) across settings with varying numbers of \rl{}s. In each setting, five agents compete: the number of \rl{}s ranges from one to five, and the remaining agents are \prompt{}s. Each setting with $j$ \rl{}s is abbreviated as $j$ \textbf{RA agent}.}
    \label{fig_multirl_5}
\end{figure}

\section{Relevance Judgments}\label{app:annotation}

\begin{table}[t]
\centering
\caption{Mean relevance judgment per configuration, player, and round. $r$ denotes statistical significance difference between rounds ($01$ vs.$30$) for the same agent and setting. 
$p$ denotes statistical significance difference between the NA and \rl{}s at the same round and setting. \textbf{Mean Rel.} is the mean relevance of the documents in the respective configuration. $\mathbf{\kappa}$ is the inter-annotator agreement rates (free-marginal multi-rater Kappa) of the relevance judgment.}

\begin{tabular}{l|cc|cc|c}
\toprule
\multirow{2}{*}{Configuration} & \multicolumn{2}{c|}{\prompt{}} & \multicolumn{2}{c|}{\rl{}} & \multirow{2}{*}{Mean Rel. / $\kappa$} \\
\cmidrule(lr){2-3} \cmidrule(lr){4-5}
& Round $01$ & Round $30$ & Round $01$ & Round $30$ & \\
\midrule
\nivList{}+\staticDocsGeneration{}+\hetroEval{} & $1.83$ & $2.80^{r}$ & $2.73^{p}$ & $2.97$ & $2.58$ / $79\%$ \\
\nivList{}+\dynamicDocsGeneration{}+\homoEval{}  & $1.83$ & $2.30$ & $2.73^{p}$ & $2.23$ & $2.27$ / $54\%$ \\
\nivList{}+\dynamicDocsGeneration{}+\hetroEval{} & $1.83$ & $2.77^{r}$ & $2.73^{p}$ & $2.80$ & $2.53$ / $76\%$ \\
\bottomrule
\end{tabular}
\label{tab:annotate-R}
\end{table}

We annotated the datasets corresponding to competitions with Mistral agent: \nivList{}+\staticDocsGeneration{}+\hetroEval{}, \nivList{}+\dynamicDocsGeneration{}+\homoEval{}, \nivList{}+\dynamicDocsGeneration{}+\hetroEval{}. Each document was judged for binary relevance to a query by three crowd workers (English speakers) on the Connect platform via CloudResearch \citep{noauthor_introducing_2024}. We adopted the annotation guidelines from MS MARCO \citep{Bajaj2016Msmarco,craswell_overview_2025}. The final relevance grade was defined as the number of annotators who marked it as relevant.

Due to budget limitations, we annotated only the \rl{} and one (arbitrarily chosen) \prompt{} for rounds 1 and 30, enabling us to analyze the effect of the alignment process (i.e documents in round $1$) and the competition dynamics (round $30$). The inter-annotator agreement, measured with the free-marginal multi-rater Kappa statistic \citep{fleiss_measuring_1971}. The kappa agreement for the relevance judgments ranged between $54\%$–$79\%$.

We observe a clear distinction between the RA and the \prompt{}s at the beginning of the competition. In round $1$, the \rl{} produces documents with higher average relevance ($2.73$ for \rl{}s in all three settings vs. $1.83$ for the \prompt{}), which we attribute to the alignment process during training that directly optimizes for ranker-preferred modifications. By round 30, however, this advantage diminishes, reflecting the herding effect whereby all agents progressively adapt toward the same high-relevance regions of the document space. When analyzing results per agent, we find that participation in the competition improves the relevance of the \prompt{}. For the \rl{} a minor improvement was observed for the \hetroEval{} settings.

\section{Declaration of Generative AI Usage in the Writing Process}
We used an LLM (OpenAI’s GPT-5) as a general-purpose writing assistant to improve the clarity and style of the paper. Its role was limited to language refinement and formatting support; all research ideas, methods, experiments, and analyses were carried out by the authors.






\end{document}